\documentclass[twocolumn]{aastex61}
\usepackage{color}

\newcommand\aastex{AAS\TeX}

\def\mdot{\hbox{M$_{\odot}$}} 
\def\coreno{\hbox{$413$}}
\def\totalno{\hbox{$512$}}

\shorttitle{\aastex\ 15 pc M Dwarfs With Masses $0.1 \leq M/\mdot \leq 0.3$}
\shortauthors{Winters et al.}

\begin{document}

\title{The Volume-Complete Sample of M Dwarfs with Masses $0.1 \leq M/\mdot \leq 0.3$ ~within 15 Parsecs}

\correspondingauthor{Jennifer G. Winters}
\email{jennifer.winters@cfa.harvard.edu}

\author[0000-0001-6031-9513]{Jennifer G. Winters}
\affil{Center for Astrophysics $\vert$ Harvard \& Smithsonian, 60 Garden Street, Cambridge, MA 02138, USA}

\author[0000-0002-9003-484X]{David Charbonneau}
\affil{Center for Astrophysics $\vert$ Harvard \& Smithsonian, 60 Garden Street, Cambridge, MA 02138, USA}
  
\author[0000-0002-9061-2865]{Todd J. Henry}
\affil{RECONS Institute, Chambersburg, Pennsylvania, 17201, USA}

\author{Jonathan Irwin}
\affil{Center for Astrophysics $\vert$ Harvard \& Smithsonian, 60 Garden Street, Cambridge, MA 02138, USA}

\author[0000-0003-0193-2187]{Wei-Chun Jao}
\affil{Department of Physics and Astronomy, Georgia State University, Atlanta, GA 30302-4106, USA}

\author[0000-0003-1645-8596]{Adric R. Riedel}
\affil{Space Telescope Science Institute, Baltimore, MD 21218, USA}

\author{Kenneth Slatten}
\affil{RECONS Institute, Chambersburg, Pennsylvania, 17201, USA}

\begin{abstract}

M dwarfs with masses $0.1 \leq M/\mdot \leq 0.3$ are under increasing scrutiny because these fully convective stars pose interesting astrophysical questions regarding their magnetic activity and angular momentum history. They also afford the most accessible near-future opportunity to study the atmospheres of terrestrial planets. Because they are intrinsically low in luminosity, the identification of the nearest examples of these M dwarfs is essential for progress. We present the  volume-complete, all-sky list of \totalno ~M dwarfs with masses $0.1 \leq M/\mdot \leq 0.3$ and with trigonometric distances placing them within 15 pc ($\pi_{trig}$ $\geq$ 66.67 mas) from which we have created a sample of \coreno ~M dwarfs for spectroscopic study. We present the mass function for these \totalno ~M dwarfs, which increases with decreasing stellar mass in linear mass space, but is flat in logarithmic mass space. As part of this sample, we present new $V_{J}R_{KC}I_{KC}$ photometry for 17 targets, measured as a result of the RECONS group's long-term work at the CTIO/SMARTS 0.9m telescope. We also note the details of targets that are known to be members of multiple systems and find a preliminary multiplicity rate of $21\pm2$\% for the primary M dwarfs in our sample, when considering known stellar and brown dwarf companions at all separations from their primaries. We further find that $43\pm2$\% of all M dwarfs with masses $0.1 \leq M/\mdot \leq 0.3$ are found in multiple systems with primary stars of all masses within 15 pc. 


\end{abstract}



\section{Introduction} 
\label{sec:intro}

Nearby stars provide the best samples of stars for study. First, nearby stars afford the brightest examples of any type, which is particularly valuable for those classes of stars with intrinsically low luminosities, such as M dwarfs with masses $0.1 \leq M/\mdot \leq 0.3$ (corresponding roughly to spectral types M4~V $-$ M7~V). Much remains unknown about these fully convective stars, such as their magnetic activity, angular momentum history, multiplicity, ages, and sizes. Thus, the nearest examples offer perhaps the best opportunity for progress in understanding the physics of this population as a whole. Second, their proximities allow angular resolution to translate into finer physical resolution than for more distant objects, permitting deeper probes for companions and direct measurements of stellar radii. In addition to the value of these targets for understanding stellar astrophysics, results have shown that it is only the terrestrial planets transiting the nearest M dwarfs with radii $<$0.3 R/R$_{\odot}$ that will be accessible for atmospheric studies with large telescopes in the near future \citep{Charbonneau(2007),Snellen(2013),Rodler(2014),Morley(2017),Lopez-Morales(2019)}. Our knowledge of these precious planets depends critically upon understanding their faint host stars. 

Even for the nearest stars, the creation of a census has historically been challenging, resulting in many samples that are biased because they are magnitude-limited. Volume-limited samples ameliorate many of the biases inherent to magnitude-limited samples, but all targets require accurate distances, which are only available via trigonometric parallaxes. 







Trigonometric parallaxes have been measured from the ground for the very nearest M dwarfs for decades  \citep{vanAltena(1995),Gatewood(2003),Costa(2005),Henry(2006),Smart(2010b),Khovritchev(2013),Dahn(2017),Henry(2018)}. However, robust results require significant time investment, so not all nearby stars have parallax measurements. The typical sequence of adding a target to an astrometry program for a parallax measurement is a long one. Searches are conducted for stars with large proper motions that indicate objects that are potentially nearby. The next step is to estimate the distances of the nearby candidates via spectroscopy and/or photometry. Only after an object has shown some evidence that it is nearby is it added to a program. It is then usually a number of years before a measurement is considered robust. Due to the low proper motion and/or 
faintness of a particular target (which requires significant telescope investment), some nearby stars have been missed or purposely omitted from target lists; thus, they lack a parallax measurement. 

Astrometry measured from space is generally superior to that measured from the ground. The astrometric satellite mission $Hipparcos$ measured parallaxes for more than 100,000 stars, but was magnitude-limited, and thus provided very few new parallax measurements for the nearest M dwarfs with masses within our mass range of interest (\citealt{Perryman(1997)}, updated in \citealt{vanLeeuwen(2007)}). The {\it Gaia} satellite \citep{Gaia(2016a)} provided an unprecedented number of trigonometric parallaxes for astronomical objects in its second data release \citep[DR2;][]{GaiaDR2(2018),Lindegren(2018)}. But because it is an ongoing survey, it is still incomplete, even for the nearest stars. This is a known phenomenon \citep{Arenou(2018), Winters(2019a)} and is due to nearby stars' typically large proper motions and/or due to unresolved binary star systems that make fitting a parallax-only model to the data difficult, at least for the first 22 months of data included in the DR2.

By augmenting the precise parallaxes from the {\it Gaia} DR2 with those from long-term, ground-based astrometry programs and $Hipparcos$, it is now possible to create volume-complete or nearly volume-complete samples for some types of stars within carefully defined distance horizons.

While the nearest 5 or 10 pc volumes are conventional distance horizon limits \citep{vandeKamp(1971),Henry(2006),Henry(2018)}{\footnote{updates found at \url{recons.org}}} for nearby star studies, samples encompassing larger volumes provide a basis for more robust statistics for various astrophysical studies. If we are to study the atmospheres of terrestrial planets, they need to orbit small stars that are nearby. Studies have shown that spaced-based missions such as the {\it James Webb Space Telescope} and upcoming giant, ground-based, telescopes such as the European Extremely Large Telescope, the Thirty Meter Telescope, or the Giant Magellan Telescope require stars smaller than 0.3 R/R$_{\odot}$ and closer than 15 pc. However, M dwarfs less massive than 0.1 M/\mdot ~that are beyond even 10 pc are too faint ($R > 15$ mag) to achieve the signal-to-noise ratios needed for high-resolution spectroscopic work on 1.5-m telescopes without exposure times that are prohibitively long (longer than one-hour). Thus, we choose 15 pc as our distance horizon and choose M dwarfs within the mass range $0.1 \leq M/\mdot \leq 0.3$ ~as our sample. We expect that roughly 368 M dwarf primaries\footnote{ We use {\it primary} to denote either a single star that is not currently known to have a companion or the most massive (or brightest in $V$) component in a multiple system; we use {\it companion} to refer to a physical member of a multiple
system that is less massive (or fainter, again in $V$) than the
primary star. } with masses $0.1 \leq M/\mdot \leq 0.3$ ~lie within 15 pc, extrapolating from the 109 found within 10 pc \citep[from][]{Winters(2019a)} and assuming a constant stellar density. Here we present our volume-complete list of M dwarfs with masses $0.1 \leq M/\mdot \leq 0.3$ ~within 15 pc.


Since September 2016, several of us (JW, DC, and JI) have been conducting an all-sky, multi-epoch, high-resolution spectroscopic survey of all known M dwarfs with masses $0.1 \leq M/\mdot \leq 0.3$ ~within 15 pc. For targets north of $\delta$ $=-$15\arcdeg, we are using the Tillinghast Reflector Echelle Spectrograph (TRES; $R \approx 44,000$) on the 1.5m telescope at the Fred Lawrence Whipple Observatory (FLWO) on Mt. Hopkins, AZ. For targets south of $\delta =-$15\arcdeg, we are using the CTIO HIgh ResolutiON (CHIRON; $R \approx 80,000$ via slicer mode) spectrograph at the Cerro Tololo Inter-American Observatory / Small and Moderate Aperture Research Telescope System (CTIO / SMARTS) 1.5m telescope. 

The goals of our survey are numerous: measure multi-epoch radial (RV) and rotational velocities ($v \sin i$); identify binaries and, for the ones with periods less than 3 years, characterize their orbits; and measure the equivalent widths of chromospheric activity indicators, including H${\alpha}$. Previous papers presented spectroscopic orbits for some of the binaries in our sample \citep{Winters(2018), Winters(2020a)}. Future papers from this project will present the radial and rotational velocities and $UVW$ space motions of these stars, as well as an analysis of their multiplicity characteristics. Others in our group are combining the spectroscopic results with $TESS$ data on these stars, with which we are determining their flare rates and photometric rotation periods \citep{Medina(2020)}, and are determining the occurrence rates of massive planetary companions around these stars. Finally, we are gathering high-resolution speckle imaging data for the sample, which will complete the coverage for companions at all separation regimes for this important population of stars. 

We began acquiring data in September 2016, but revised our target list in 2018 with the availability of the second data release from {\it Gaia}. Thus, we describe the history of the sample in \S \ref{sec:sample_create} and the characteristics of our sample in \S \ref{sec:sample}. We then discuss in \S \ref{sec:disc} the completeness and biases of the sample presented.

\section{Creation of the Sample} \label{sec:sample_create}

Our goal was to compile a complete sample of M dwarfs with masses $0.1 \leq M/\mdot \leq 0.3$ ~within 15 pc for our spectroscopic survey. Creating this list requires knowledge of the stars' distances and masses. For nearby stars with small parallax uncertainties, the trigonometric parallax provides their distances.\footnote{For a description of calculating distances from parallaxes for non-nearby stars, see \citet{Luri(2018)}.} But because we can only directly measure masses for stars in binary systems, we rely on the Mass-Luminosity Relation (MLR) to estimate the masses of single stars. As we describe below, we accomplish this through the use of 2MASS $K_{s}-$band photometry. A third, more subtle constraint on sample membership is the consideration of close binaries in the sample. 

\subsection{Initial Sample Selection}
\label{subsec:sample_initial}

Our initial sample was created in 2016. We first identified all types of M dwarfs (main sequence stars with estimated masses 0.075 -- 0.64 M/\mdot) that had trigonometric distances placing them within 15 pc ($\pi_{trig}$ $\geq$ 66.67 mas), with no constraint on the parallax uncertainties. We began with the volume-limited, all-sky sample of 1120 M dwarfs presented in \citet{Winters(2019a)}\footnote{We used the target list  while the paper was in preparation.} that contains many M dwarfs with mass estimates that lie within 25 pc via a trigonometric parallax; however, that sample was finalized as of 01 January 2014, and thus did not include M dwarfs with parallaxes that were subsequently published. We then identified M dwarfs that had new parallaxes placing them within 15 pc from more recent parallax papers \citep{Dittmann(2014),Weinberger(2016),Finch(2016)}, which we added to our list. As new parallaxes for nearby M dwarfs were published \citep[e.g., by][]{Dahn(2017),Finch(2018)}, we added these to our list. 



We then determined which targets fell into our mass region of interest. We extracted $JHK_{s}$ magnitudes from the 2MASS Point Source Catalog \citep{Cutri(2003),Skrutskie(2006)}, which we confirmed by eye to correspond to the star in question. We calculated the masses of all the M dwarfs using the mass-luminosity relation by \citet{Benedict(2016)} with the assumption that they were single, main-sequence stars. We used the absolute $K_{s}-$band magnitude for the mass estimation because it is known to be less sensitive to metallicity effects \citep{Henry(1993),Delfosse(2000),Benedict(2016)}. In the cases where trigonometric parallaxes were not available\footnote{This was the case only for the initial sample creation. As we note below in \S \ref{subsec:sample_revised}, the DR2 provided parallaxes for all stars that previously had only photometric distance estimates.}, photometric distance estimates from \citet{Winters(2015)} were used for the mass estimate.

Known binaries with blended $K_{s}-$band photometry required special treatment to determine the primary stars' masses. These fell into four categories: 1) binaries with published orbital solutions that reported masses for the individual components; 2) binaries with published orbital solutions that reported mass ratios, but not individual masses (i.e., spectroscopic binaries); 3)  binaries with high-resolution imaging results that reported a magnitude difference ($\Delta m$) between the components; 4) astrometric binaries with no reported $\Delta m$.

For the multiple systems with orbital solutions that reported masses (category 1), we simply adopted these masses. For binaries with published orbital solutions for double-lined spectroscopic binaries that offered mass ratios but no reported masses (category 2), we estimated the masses of the component stars in the same way as presented in \citet{Winters(2019a)}. We describe it here. We first estimated a $\Delta K_{s}$ magnitude based on an assumed flux ratio between the components. We then deblended the $K_{s}-$band photometry and calculated the components' M$_{K_{s}}$. We then estimated the components' masses using the MLR and then calculated the mass ratio for the system. We repeated these steps after adjusting the $\Delta K_{s}$ until our estimated mass ratio agreed with the mass ratio in the published orbital solution. As a check that this method results in accurately estimated masses, we calculated the masses of the binary star components in category 1 the same way. Overall, we find a standard deviation of 0.013 M/\mdot when comparing the published masses to our estimated masses for the binary systems in category 1, in agreement with the scatter of 0.014 M/\mdot ~in the MLR, as reported in \citet{Benedict(2016)}. We did not include GJ~866ABC as part of this check because the photometry includes three unresolved stars, making the test more complicated. When checking the results for GJ~791.2AB, we found discrepancies of 0.083 M/\mdot ~and 0.014 M/\mdot ~for the primary and secondary components, respectively, which we attribute to the DR2 parallax. There is a difference of 20 mas between the DR2 and weighted mean literature parallax values (see Figure \ref{fig:pi_diff}). It also has large values for the DR2 astrometric  flags `astrometric excess noise' (6.71 mas, the highest point in Figure \ref{fig:ast_noise}) and its significance (20300). Therefore, we used the ground-based parallax for GJ~791.2AB for our check and find differences of 0.050 M/\mdot ~and 0.004 M/\mdot ~for the primary and secondary star masses. The discrepancy between our calculated and the published mass for GJ~791.2A is by far the largest of the 28 components' masses we checked. We therefore conclude that our method is robust. 

For reported binaries with blended $K_{s}-$band photometry but without published orbits (category 3), we deblended the photometry based on delta-magnitude ($\Delta m$) information available in the literature from high-resolution imaging studies (e.g., adaptive optics, speckle, lucky imaging work) before calculating the mass of the primary star. In the cases where the $\Delta m$ was measured in a filter that was not $K_{s}$, we converted the $\Delta m$ to $\Delta K_{s}$ using the transformations in \citet{Riedel(2014)}. 

Six systems are astrometric binaries reported in publications by the REsearch Consortium On Nearby Stars (RECONS)\footnote{\url{www.recons.org}} group, which had no reported $\Delta m$ information (category 4). For these systems, we assumed all companions were stellar in nature, unless otherwise stated, and we follow the same procedure for astrometric binaries as in \citet{Winters(2019a)}, which we describe here. We calculated the photometric distance estimates for these six systems using the relations from \citet{Henry(2004)}. All of the photometric distances were in agreement with or smaller than the trigonometric distances. We then calculated the ratio of the trigonometric distance to the photometric distance, which we used to estimate the $\Delta m$ between the components. We note that an equal luminosity binary, with $\Delta m$ near zero, will not be detected astrometrically because the photocenter will not be perturbed.  For systems where the ratio was $\leq 1.3$, we adopted a $\Delta m$ of 3 mag between the components, as the agreement between the two distances indicates that the companion contributes a negligible amount of light to the system's photometry. For systems where the ratio was $> 1.3$, indicating a luminous companion which contributes light to the system's photometry, we adopted a $\Delta m$ of 2 mag between the components. Because of the 15\% uncertainties on the photometric distances, it is difficult to place tighter constraints on the distance ratio. Thus, we are unable to fine-tune the magnitude differences until we acquire high-resolution imaging of these astrometric binaries. The adopted magnitude difference was in the filter in which the parallax was measured, which we then converted to $\Delta K_{s}$, again using the transformations in \citet{Riedel(2014)}. Finally, we selected all stars with masses $0.1 < M/\mdot < 0.3$, regardless of the mass uncertainty. 



\subsection{Sample Revision with DR2}
\label{subsec:sample_revised}

With the availability in 2018 of parallaxes in the {\it Gaia} second data release (DR2) \citep{GaiaDR2(2018),Lindegren(2018)}, we revised our sample. We first searched the DR2 via {\it Vizier} for all objects with parallaxes larger than 66.667 mas (distances nearer than 15 pc) with no astrometric quality cuts and no constraint on parallax uncertainty. Using their proper motions, we adjusted the coordinates of the stars in our initial sample from J2000.0 to J2015.5, the epoch of the DR2 positions, which we then cross-matched against the DR2 15 pc list. The {\it Gaia} archive cross-match to the 2MASS Point Source Catalog \citep{Cutri(2003),Skrutskie(2006)} provides only a 2MASS identifier, so 2MASS $K_{s}$ magnitudes were not readily available for mass calculations. To determine which were M dwarfs with masses within our mass range of interest, we used the presumed single stars in our initial list to determine the typical ($B_p-R_p$) colors and $G$ ~magnitudes for these M dwarfs. We then imposed color limits of $2.3 < (B_p-R_p) < 4.5$ mag to identify nearby M dwarfs with masses within our range of interest that previously did not have published parallaxes that placed them within 15 pc. We also only chose stars with $G <$ 16 mag, as the least massive star LHS~1604, with mass 0.10 M/\mdot ~at roughly 15 pc has $G = $ 15.34 mag. This apparent magnitude cut also served to eliminate any false stars or artefacts, as suggested by \citet{Lindegren(2018)} and \citet{Arenou(2018)}. Finally, we extracted 2MASS $JHK_{s}$ magnitudes for the newly identified nearby stars and recalculated masses using the DR2 parallaxes, where available, for the entire list because a revision in the parallax will affect the mass estimate and thus the sample membership. Again, we selected all stars with masses $0.1 < M/\mdot < 0.3$, regardless of the mass uncertainty.


\subsection{Sample Members That Are Secondaries}
\label{subsec:secondaries}

We did not restrict our search to primary stars alone. We identified a total of 143 secondaries with masses within our mass range of interest, as outlined in Table \ref{tab:all_midMs}. Sixty-seven of these M dwarfs are companions to M dwarf primaries with masses $0.1 < M/\mdot < 0.3$ within 15 pc. Ten companions are widely separated ($>$ 4\arcsec) from their primaries and are included explicitly in our spectroscopic sample. The remaining 55 are at separations $<$ 4\arcsec ~from their primaries and are thus implicitly included in our sample, along with 28 companions that are less massive than 0.1 M/\mdot ~(all separated by $<$4\arcsec ~from their primaries). Included in the 143 are three M dwarf pairs (GJ~376C, LP~993-116B and GJ~630.1B) that are members of hierarchical triples with massive primaries with separations $<$4\arcsec ~between the pair. 

The remaining 73 M dwarf companions with masses $0.1 \leq M/\mdot \leq 0.3$ ~we identified during our search of more massive stars within 15 pc. From the sample in \citet{Winters(2019a)}, we searched the early M dwarf primaries (masses $0.30 < M/\mdot \leq 0.64$) to identify those with M dwarf companions. We then used the RECONS 25 pc database to identify M dwarf companions to primaries that were not M dwarfs (i.e., A,F,G,K, and white dwarfs)\footnote{We consider the M dwarf to be the secondary in the cases of M dwarf - white dwarf systems, as the white dwarf was previously the more massive, and thus primary, component in the system.} within 15 pc. We searched the literature and flagged as M dwarf companions those that had reported masses or estimated spectral types within our range of interest. In a few cases, magnitude differences were reported, from which we estimated the component's mass using the MLR. There are 33 M dwarfs with masses $0.1 \leq M/\mdot \leq 0.3$ that are companions to AFGK, early M, and white dwarf primaries, which are separated by more than 4\arcsec ~from their primaries. These are all included in our spectroscopic sample because they are not contaminated by light from the more massive components. We additionally include the one M dwarf (G~203-47) that is a close companion to a white dwarf in a 15.7-day orbital period \citep{Delfosse(1999c)}; the white dwarf adds a negligible amount of light to the system at 7100\AA, the part of the spectrum that we use for our RV measurements. The remaining 39 of the identified M dwarf companions have angular separations $<$4\arcsec ~from their more massive primaries. One additional unseen component of a single-lined spectroscopic binary (GJ~92B) has only a mass range reported, meaning that the secondary could very well be an M dwarf with a mass in our range of interest, but more data are needed to constrain its mass.

\begin{figure}
\includegraphics[scale=.42,angle=0]{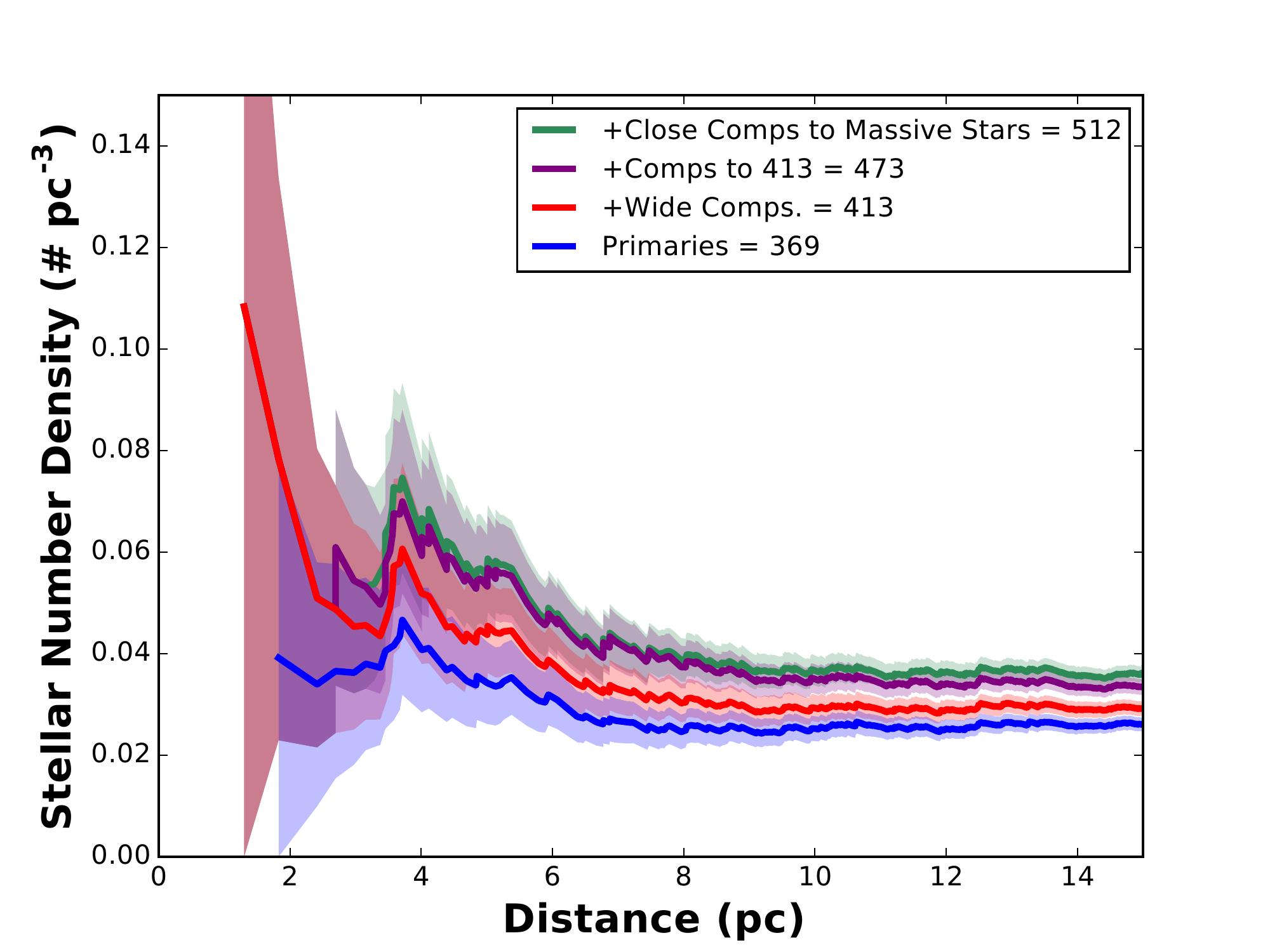}
\caption{The number of M dwarfs with masses $0.1 \leq M/\mdot \leq 0.3$ ~per cubic parsec as a function of distance. In blue are noted the 369 primaries; the red line indicates the \coreno ~in our spectroscopic sample (369 primaries plus 44 companions); the purple line includes the 60 close M dwarf companions (separations $<$ 4\arcsec) to the \coreno ~stars in our core sample; in green are all known M dwarfs with masses $0.1 \leq M/\mdot \leq 0.3$ ~within 15 pc, which includes the 39 close companions to more massive stars. Illustrated is the fairly constant density from 8-15 pc, once small number statistics are overcome at the closest horizons.  This implies the sample is volume-complete. Uncertainty bands indicate the Poisson errors on the number of stars in each bin.  \label{fig:cum_hist}}
\end{figure}

We list in Table \ref{tab:all_midMs} the distributions of the known M dwarfs with masses $0.1 \leq M/\mdot \leq 0.3$ ~within 15 pc. We show the running unbinned stellar density of these M dwarfs within 15 pc in Figure \ref{fig:cum_hist}, broken into four subsets: 1) 369 presumed single and primary stars; 2) subset one, plus the addition of 43 companions at separations $>$4\arcsec ~from their primaries, plus G~203-47; 3) subset two, plus the addition of unresolved companions to subset two stars; and 4) subset three, plus the addition of unresolved companions to massive stars. Each bump in each curve indicates the addition of a new M dwarf within our selected mass range. Also highlighted is that the nearest primary (Barnard's Star) is more distant than than the nearest of these M dwarfs (Proxima Centauri), a companion to a more massive primary. Illustrated is a fairly uniform stellar density from 8-15 pc, once small number statistics are overcome at the closest horizons. This implies the sample is volume-complete.

\begin{deluxetable*}{lccc}
\tabletypesize{\small}
\tablecaption{Known M Dwarfs With Masses $0.1 \leq M/\mdot \leq 0.3$ ~Within 15 pc \label{tab:all_midMs}}
\tablecolumns{4}
\tablenum{1}

\tablehead{
\colhead{Type of Star}      & 
\colhead{Sep $<$4\arcsec}         &
\colhead{Sep $>$4\arcsec}         &
\colhead{Total}
}

\startdata
Single Mid-M Dwarf                     &    &     & {\bf 290} \\
Mid-M Dwarf Primary                    &    &     & {\bf 79} \\
Comp. to Mid-M Dwarf Primary           & {\it 57} & {\bf 10}  & 67  \\
Comp. to Mid-M Dwarf Comp.             &  {\it 3} &  0  &  3\\
Comp. to Early M Dwarf Primary         & 26 & {\bf 16}  & 42  \\       
Comp. to K Dwarf Primary               &  5 &  {\bf 7}  & 12   \\  
Comp. to G Dwarf Primary               &  8 &  {\bf 3}  & 11  \\  
Comp. to F Dwarf Primary               &  0 &  {\bf 1}  &  1  \\ 
Comp. to A Dwarf Primary               &  0 &  {\bf 1}  &  1 \\  
Comp. to White Dwarf Primary           &  {\bf 1} &  {\bf 5}  &  6   \\ 
\hline
                                       & 100  & 43  & 512 \\ 
\hline
\enddata
\tablecomments{In bold are noted the numbers of stars from each type that are included in our spectroscopic sample of \coreno ~M dwarfs. In italics are noted the 60 mid-M dwarfs that are implicitly included as unresolved companions in the spectroscopic sample \coreno ~M dwarfs.}
\end{deluxetable*}

\begin{deluxetable}{lccc}
\tabletypesize{\small}
\tablecaption{Parameters of the M Dwarfs Presented in this Paper \label{tab:params}}
\tablecolumns{2}
\tablenum{2}

\tablehead{
\colhead{Parameter}      & 
\colhead{Value}          
}

\startdata
Mass (M/\mdot)       & $0.1 \leq M/\mdot \leq 0.3$      \\
$M_{G}$ (mag)         & $10.4 \leq M_{G} \leq 14.5$   \\
($B_{p}-R_{p}$) (mag)   & $2.3 \leq (B_{p}-R_{p}) \leq 4.5$     \\
$M_{K}$ (mag)         & $7.0 \leq M_{K} \leq 9.5$     \\
($R_{KC}-K_{S}$) (mag)  & $3.1 \leq (R_{KC}-K_{S}) \leq 6.0$     \\       
\enddata
\end{deluxetable}

\subsection{Possibly Hidden Sample Members}
\label{subsec:poss_hidden}

Because one of the goals of our survey is a search for companions, we wanted to gather data for presumably single M dwarfs that had estimated masses larger than our upper mass limit of $0.3$ M/M$_{\odot}$, but which would result in stellar components with masses within our mass range, should they be discovered to host equal or nearly-equal mass companions. An unresolved, equal-luminosity M dwarf binary where both components have masses of $0.3$ M/M$_{\odot}$ will have an estimated total mass of $0.44$ M/\mdot ~from its blended M$_{K_{s}}$ photometry. These types of systems would be distinguishable by their overluminosity. We identified from the master M dwarf list described in \S \ref{subsec:sample_initial} 18 stars with estimated masses $0.30-0.44$ M/M$_{\odot}$, which were overluminous and for which we are gathering reconnaissance spectroscopy. We excluded targets in this mass range that had both high-resolution imaging and spectroscopy results in the literature, under the assumption that those programs would have detected a companion. We further include in this list of candidates four stars that have estimated masses $0.300-0.304$ M/\mdot ~and two M dwarfs just beyond the 15 pc horizon with large parallax uncertainties. 

It is possible that there are M dwarf companions to AFGKs that have not yet been identified. In fact, three candidate members have been noted recently as unresolved companions to nearby K dwarfs (D.Nusdeo, in review), but which need follow-up observations for confirmation. Should any new M dwarfs within our mass range of interest be detected, we will take note, but they will likely not become part of our spectroscopic survey due to their proximity to their bright primary stars.

\section{The Sample} \label{sec:sample}

Of all the M dwarfs described above, the final sample for our spectroscopic survey comprises \coreno ~M dwarfs with masses $0.1 \leq M/\mdot \leq 0.3$ ~within 15 pc. All have trigonometric parallaxes. {\bf These \coreno ~include} 292 presumably single M dwarfs, 77 M dwarf primaries of multiple systems, {\bf 43 companions that are separated} by more than 4\arcsec ~from their primary (of any mass), and one close companion to a faint white dwarf. In Table \ref{tab:all_midMs}, these numbers are in bold print. We refer to this as our `spectroscopic' sample for the remainder of the paper. We list the range of masses for these kinds of stars, as well as some of their observational parameters in Table \ref{tab:params}. In Table \ref{tab:sample}, we present the relevant astrometry, photometry, and multiplicity information for all \totalno ~M dwarfs with masses $0.1 \leq M/\mdot \leq 0.3$ ~within 15 pc. The \coreno ~stars in our spectroscopic survey are indicated with the sample code `s', their 60 close companions are noted with code `sc', and the 39 close companions to more massive stars are noted with code `mc'. Although the 39 companions to more massive stars are M dwarfs within 15 pc within our mass range of interest (or a candidate, as in the case of GJ~92B), we do not include them in our spectroscopic survey sample because their separations are too small to avoid contamination from their bright primaries. The DR2 reports parallaxes for 388 of the \coreno ~stars in our spectroscopic sample, while accurate parallaxes existed for 394 of these \coreno ~before the availability of the DR2.

In Table \ref{tab:rejected}, we list 120 stars in the same format as Table \ref{tab:sample}. These include the 95 that we excluded from our spectroscopic sample due to mass or distance revisions with the availability of the DR2 parallaxes, and the 25 candidate stars. Some of these 25 could move into our spectroscopic sample, should they be revealed to be equal-luminosity binaries.

\subsection{Astrometry}

For our spectroscopic sample membership determination and for all analyses, we adopt only the DR2 parallaxes, where available; otherwise, we use the weighted mean literature values. Parallaxes that were measured for these nearby M dwarfs before the availability of the {\it Gaia} DR2 come from myriad sources. The largest source is the {\it General Catalogue of Trigonometric Parallaxes, Fourth Edition} by Yale University Observatory, usually called the Yale Parallax Catalog \citep[YPC;][]{vanAltena(1995)}, which is a compendium of ground-based parallaxes that were published before 1995. The YPC provides trigonometric parallaxes for 182 of the M dwarfs in our spectroscopic sample, 23 of which have measurement uncertainties larger than 10 mas. RECONS has added 101 new parallaxes for these nearby M dwarfs and improved the large uncertainties on an additional 20, primarily in the southern hemisphere (\citealt{Costa(2005),Jao(2005),Jao(2011),Jao(2017),Henry(2006), Henry(2018),Riedel(2010),Riedel(2011),Riedel(2014),Riedel(2018),Dieterich(2014),Lurie(2014),Bartlett(2017),Subasavage(2017),Winters(2017),Vrijmoet(2020)}. The remaining 111 published parallaxes for the M dwarfs in our spectroscopic sample is mainly due to work by \citet{Dittmann(2014),Finch(2016),Weinberger(2016),Dahn(2017),Finch(2018)}. The $Hipparcos$ satellite provided only 6 new measurements for these stars (\citealt{Perryman(1997)}, with results updated in \citealt{vanLeeuwen(2007)}), largely due to its magnitude limit of $V \lesssim$ 12 mag.

With the release of the DR2 parallaxes, 34 M dwarfs fit our membership criteria and were thus added to our final sample. Of these, 19 (5\%) targets in our spectroscopic sample had no previously published parallax; all but two of these stars (2MA~J0049+6518 and 2MA~J0943-3833) had previously been identified as nearby candidates, but lacked a parallax. The DR2 provided improved parallaxes for fifteen additional targets that had literature parallaxes with uncertainties larger than 10 mas or parallaxes smaller than 66.67 mas. With the sample revision, 100 stars in our initial sample no longer fit the sample membership criteria and were removed from our spectroscopic sample due to distance or mass revisions. 


In our spectroscopic sample, 25 ($6$\%) of the stars do not have {\it Gaia} DR2 parallaxes, but have published ground-based parallaxes. This is in agreement with the 6\% of the volume-limited sample of 1120 M dwarfs within 25 pc that lack parallaxes in the DR2, as noted in \citet{Winters(2019a)}. In some cases, the typically large proper motions of nearby stars make centroiding challenging \citep{Arenou(2018)}. In other cases, they are binaries with companions at sub-arcsecond separations that also make it hard to determine the photocenter of light that is needed for the accurate measurement of a parallax \citep{Arenou(2018),Ziegler(2018)}. The DR2 only uses 22 months of data, and as yet only uses a single-star model for the astrometric solution. It is expected that data releases starting with the full DR3 will include multiplicity information. Seventeen of the 25 lacking DR2 parallaxes have only two-parameter astrometric solutions (i.e., coordinates, but no proper motion or parallax measurements). All but five of these 20 targets are known to be binaries or triples with separations $<$1\arcsec ~and $\lesssim$ 1\farcs5, respectively, between components. It is possible that these five (LEP~0617+8353, LP~549-6, LHS~296, GJ~1227, and LHS~3746) are therefore multiples; however, LHS~296 and GJ~1227 have proper motions exceeding 1\arcsec yr$^{-1}$, which could have made it difficult to fit a parallax solution for these two stars. Two of the 17 (LP~993-116(BC) and LTT~1445BC) are wide companions to more massive stars and for which we have adopted the parallax of the primary component. The remaining 8 stars do not have an entry in the DR2 (including GJ~406 $=$ Wolf 359). Six are binaries or triples with subarcsecond separations between components, while GJ~406 has a proper motion of 4\farcs7 yr$^{-1}$. GJ~618B is a wide companion whose primary has a DR2 parallax, which we have adopted. 

\begin{figure}
\hspace{-0.55cm}
\includegraphics[scale=.33,angle=270]{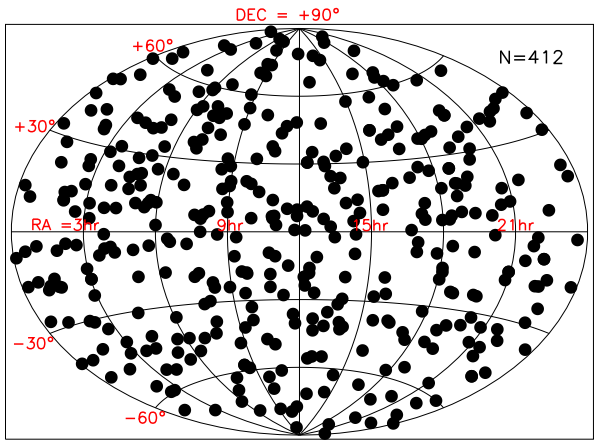}
\caption{Distribution on the sky in equatorial coordinates of the spectroscopic sample of \coreno ~nearby M dwarfs. \label{fig:aitoff}}
\end{figure}

\begin{figure}[ht!]

\hspace{-1.55cm}
\includegraphics[scale=.40,angle=90]{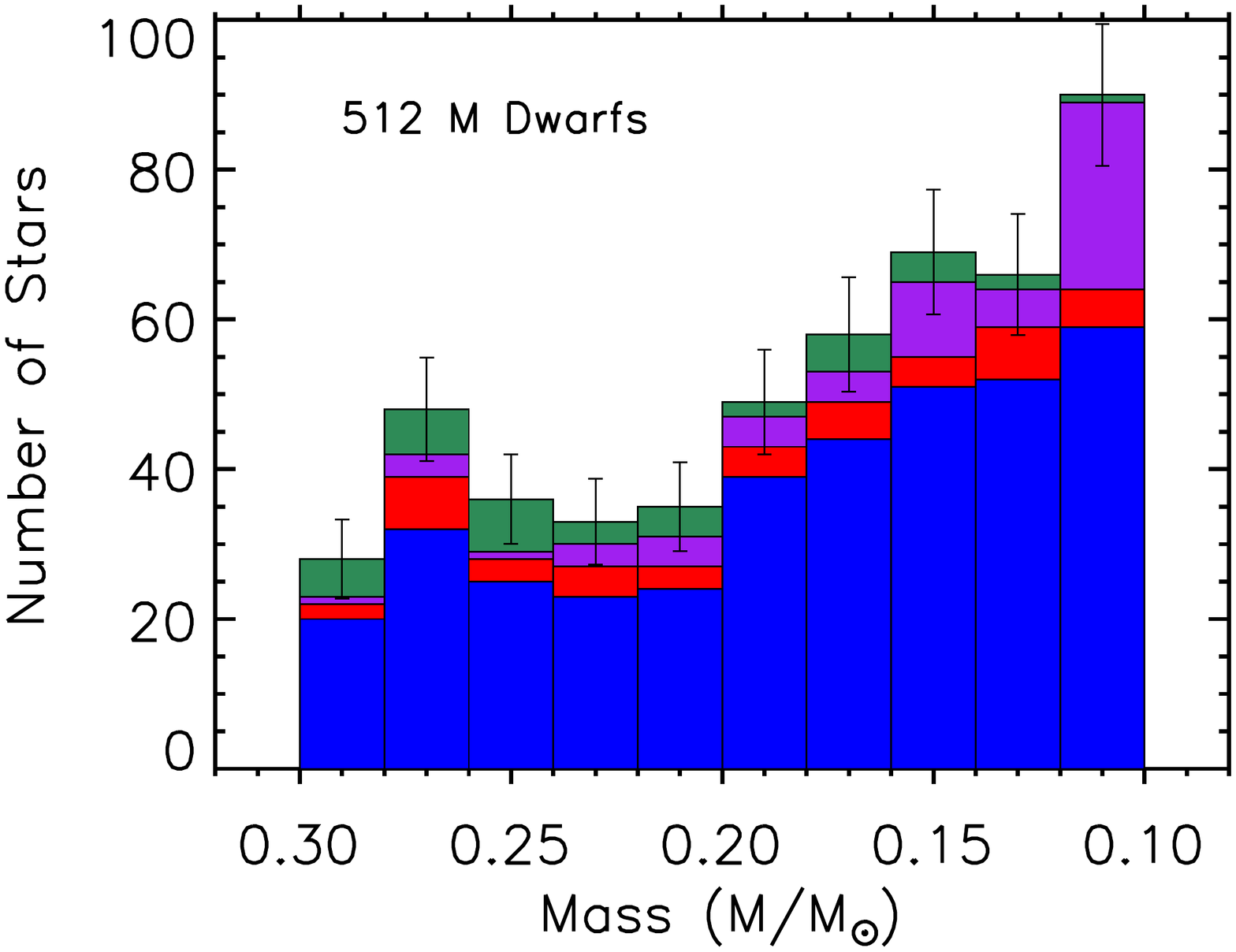}

\vspace{-1.5cm}
\hspace{-1.55cm}
\includegraphics[scale=.40,angle=90]{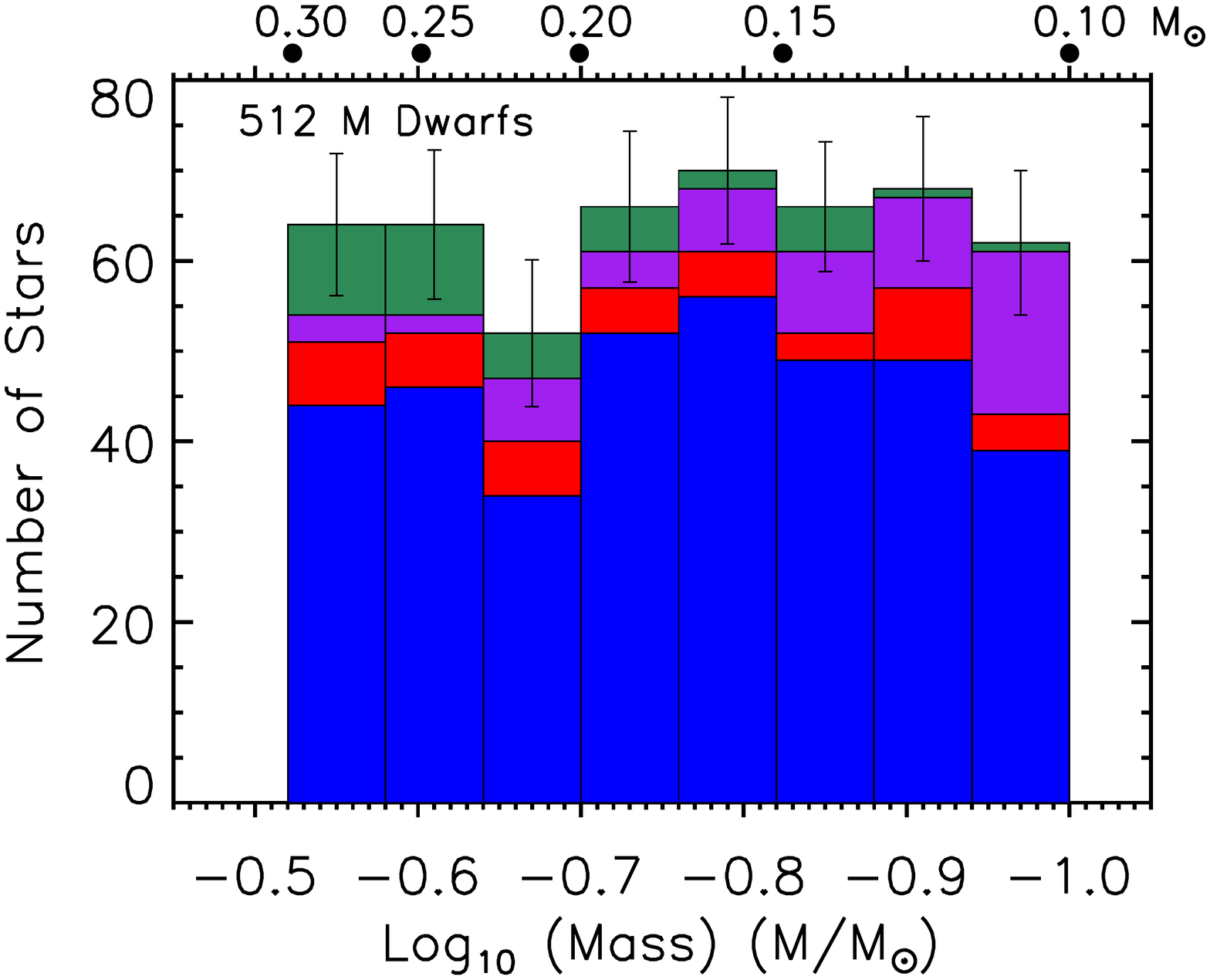}

\vspace{-1.0cm}
\caption{The mass distributions of the \totalno ~nearby M dwarfs with masses $0.1 \leq M/\mdot \leq 0.3$ ~within 15 pc, illustrated as stacked histograms in linear (top panel) and log forms (bottom panel). In blue are noted the 369 primaries; red indicates the 43 wide companions, plus G~203-47; purple indicates the 60 close M dwarf companions (separations $<$ 4\arcsec) to the \coreno ~stars in our spectroscopic sample (noted in blue and red); in green are the 39 close companions to more massive stars. The uncertainties in both panels are Poisson. We note that the bin sizes differ between the two panels. Close binaries with blended photometry were deblended before the mass estimate calculation. Masses are noted along the top of the lower panel, with points corresponding to each of the indicated masses to guide the eye. An increase in the number of objects per mass bin with decreasing mass is evident in the linear histogram, while the distribution is flat in the log version. \label{fig:mass_hist}}
\end{figure}

We show the all-sky distribution of our spectroscopic sample of \coreno ~M dwarfs in Figure \ref{fig:aitoff} in equatorial coordinates. Figure \ref{fig:mass_hist} shows the mass distribution of the different subsets of M dwarfs shown in Figure \ref{fig:cum_hist} as a stacked histogram. In addition to the distribution in linear form (top panel), we include a version in log form (bottom panel). In the top panel, we note the increase in number of stars per mass bin with decreasing stellar mass, in agreement with other mass functions for nearby M dwarfs, such as Figure 24 from \citet{Winters(2019a)}. The log version (bottom panel) is flat when including all the known primaries and companions. While other mass functions typically represent the y-axis as the logarithm of the number density per logarithmic mass bin, we do not adopt that form. However, we note that the log version of our mass function is in agreement with the single-star mass function in this mass range from \citet{Bochanski(2010)} (their Figure 23). In contrast, our mass function does not resemble Figure 2 from \citet{Chabrier(2003b)} because that work presents a system mass function, where photometry of the known components of multiple systems are blended into one system magnitude from which a mass estimate is made.

\begin{figure}
\hspace{-1.0cm}
\includegraphics[scale=.38,angle=270]{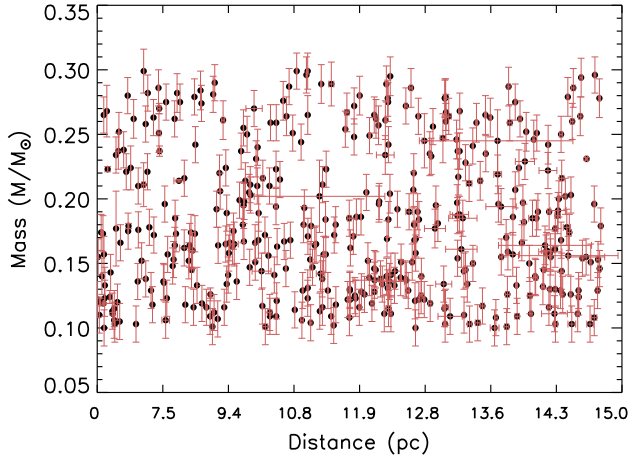}
\caption{Stellar density of the sample. Plotted is mass
versus distance in eight equal volume shells for the \coreno ~stars in our spectroscopic, all-sky sample. Close binaries with blended photometry were deblended before the mass estimate calculation. Illustrated is the fairly uniform distribution of stars to 15 pc. \label{fig:mass_density}}
\end{figure}

We showcase the volume-completeness and uniform distribution of the sample in  Figures \ref{fig:cum_hist} (discussed in \S 
\ref{subsec:secondaries}) and \ref{fig:mass_density}. In Figure \ref{fig:mass_density}, the mass uncertainties for most objects are dominated by the scatter (0.014 M/\mdot) in the MLR. The distance uncertainties are calculated to be symmetric as follows:

\begin{equation}
\sigma_{pc} = \Bigg \langle \Bigg | \left( \frac{1}{\pi} - \frac{1}{\pi \pm \sigma_{\pi}} \right) \Bigg | \Bigg \rangle
\end{equation}

\noindent A few objects without DR2 parallaxes are noticeable by their large distance uncertainties, but most objects have distance errors that are smaller than the points.

\startlongtable
\begin{deluxetable*}{clcl}
\tabletypesize{\scriptsize}
\tablecaption{Astrometry, Photometry, and Multiplicity Information for \totalno ~M Dwarfs \label{tab:sample}}
\tablenum{3}
\tablehead{\colhead{Column}          &
	   \colhead{Format}              &
	   \colhead{Units}               &
	   \colhead{Description}         
	   }
\startdata
1  & A11      & hh:mm:ss         & Right Ascension (J2000.0) \\
2  & A11      & dd:mm:ss         & Declination (J2000.0)    \\
3  & A15      &                  & Name  \\
4  & A1       &                  & Component  \\
5  & A16      &                  & 2MASS Identifier   \\
6  & A5       &                  & Configuration of Multiple Components \\
7  & A10      & arcsec           & Separation ($\rho$) Between Multiple Components  \\
8  & A3       & deg              & Position Angle ($\theta$) Between Multiple Components  \\
9  & I4       &                  & Year of Multiplicity Measurement  \\
10 & A6       &                  & Technique for Multiplicity Detection (1) \\
11 & A21      &                  & Reference for Multiplicity Detection \\
12 & F4.2     & mag              & Magnitude Difference ($\Delta m$) between Multiple Components  \\
13 & A3       &                  & Filter of Magnitude Difference Multiple Components  \\
14 & A21      &                  & Reference for Multiple Component's  Magnitude Difference \\
15 & F8.2     & mas yr$^{-1}$    & Proper Motion in RA ($\mu_{\alpha}$)   \\
16 & F8.2     & mas yr$^{-1}$    & Proper Motion in Dec ($\mu_{\delta}$)  \\
17 & A21      &                  & Reference for Proper Motion           \\
18 & F8.4     & mas              & {\it Gaia} DR2 Parallax ($\pi_{DR2}$)  \\
19 & F6.4     & mas              & Uncertainty on Gaia DR2 Parallax \\
20 & A1       &                  & Note on DR2 parallax (2) \\
21 & F5.3     & mas              & DR2 Astrometric Excess Noise   \\
22 & F8.2     &                  & Significance of DR2 Astrometric Excess Noise  \\
23 & F6.2     & mas              & Weighted Mean Literature Parallax  ($\pi_{wt}$) \\
24 & F4.2     & mas              & Uncertainty on Lit. Parallax \\
25 & A131     &                  & Reference(s) for Parallaxes included in Weighted Mean Parallax   \\
26 & F5.2     & mag              & $G$ Magnitude \\
27 & F4.2     & mag              & $(G_{B_{P}}-G_{R_{P}})$ Color  \\
28 & F5.2     & mag              & $V_{\rm J}$ Magnitude  \\
29 & F5.2     & mag              & $R_{\rm KC}$ Magnitude  \\
30 & F5.2     & mag              & $I_{\rm KC}$ Magnitude   \\
31 & A21      &                  & Reference for $V_{\rm J}R_{\rm KC}I_{\rm KC}$ Magnitudes  \\
32 & F4.1     & mag              & Transformed $R_{KC}$ Magnitude from $G$ and $K_{s}$ \\
33 & F6.3     & mag              & 2MASS $J$ Magnitude   \\
34 & F5.3     & mag              & Uncertainty on 2MASS J Magnitude \\
35 & F6.3     & mag              & 2MASS $H$ Magnitude   \\
36 & F5.3     & mag              & Uncertainty on 2MASS H Magnitude \\
37 & F6.3     & mag              & 2MASS $K_{s}$ Magnitude    \\
38 & F5.3     & mag              & Uncertainty on 2MASS $K_{s}$ Magnitude \\
39 & A1       &                  & Note on 2MASS $K_{s}$ Magnitude (3)   \\
40 & F5.3     & M/M$_{\odot}$    & Stellar Mass  \\
41 & F5.3     & M/M$_{\odot}$    & Uncertainty on Mass \\
42 & A1       &                  & Note on Mass (4)  \\
43 & A1       &                  & Note on Sample  Membership (5) \\
\enddata
\tablerefs{
This work;
\citet{Allen(2012)};
\citet{Balega(2013)};
\citet{Baroch(2018)};
\citet{Bartlett(2017)};
\citet{Benedict(1999)};
\citet{Benedict(2000b)};
\citet{Benedict(2001)};
\citet{Benedict(2016)};
\citet{Bergfors(2010)};
\citet{Bessel(1990)};
\citet{Bessell(1991)};
\citet{Beuzit(2004)};
\citet{Bowler(2015)};
\citet{Cortes-Contreras(2017)};
\citet{Costa(2005)};
\citet{Dahn(1988)};
\citet{Dahn(2017)};
\citet{Davison(2015)};
\citet{Delfosse(1999c)};
\citet{Delfosse(1999d)};
\citet{Dieterich(2012)};
\citet{Dieterich(2014)};
\citet{Dittmann(2014)};
\citet{Docobo(2006a)};
\citet{Dupuy(2017)};
\citet{Duquennoy(1988b)};
\citet{Eggenberger(2007)};
\citet{Fabricius(2000)};
\citet{Faherty(2012)};
\citet{Farihi(2005)};
\citet{Finch(2016)};
\citet{Finch(2018)};
\citet{Gatewood(2003)};
\citet{Gatewood(2008)};
\citet{Gatewood(2009)};
\citet{Golimowski(2004)};
\citet{Halbwachs(2000)};
\citet{Han(2002)};
\citet{Harrington(1981)};
\citet{Heintz(1974)};
\citet{Heintz(1988)};
\citet{Heintz(1990)};
\citet{Heintz(1991)};
\citet{Heintz(1993)};
\citet{Heintz(1994)};
\citet{Henry(1993)};
\citet{Henry(1997)};
\citet{Henry(1999)};
\citet{Henry(2004)};
\citet{Henry(2006)};
\citet{Henry(2018)};
\citet{Hershey(1998)};
\citet{Hog(2000)};
\citet{Horch(2010)};
\citet{Horch(2011a)};
\citet{Horch(2012a)};
\citet{Horch(2015)};
\citet{Hosey(2015)};
\citet{Ianna(1996)};
\citet{Janson(2014a)};
\citet{Jao(2003)};
\citet{Jao(2005)};
\citet{Jao(2011)};
\citet{Jao(2017)};
\citet{Jodar(2013)};
\citet{Khovritchev(2013)};
\citet{Koen(2002)};
\citet{Koen(2010)};
\citet{Kohler(2012)};
\citet{Lacy(1977)};
\citet{Law(2008)};
\citet{Leinert(1994)};
\citet{Lepine(2005a)};
\citet{Lepine(2009)};
\citet{Lindegren(1997)};
\citet{Lindegren(2018)};
\citet{Lowrance(2002)};
\citet{Lurie(2014)};
\citet{Luyten(1979b)};
\citet{Mamajek(2013)};
\citet{Martin(1998a)};
\citet{Martinache(2007)};
\citet{Martinache(2009)};
\citet{Mason(2018)};
\citet{Mason(2019)};
\citet{McArthur(2010)};
\citet{McCarthy(1984)};
\citet{Montagnier(2006)};
\citet{Morales(2009)};
\citet{Pravdo(2004)};
\citet{Reid(2002)};
\citet{Reid(2003)};
\citet{Reid(2004)};
\citet{Riedel(2010)};
\citet{Riedel(2011)};
\citet{Riedel(2014)};
\citet{Riedel(2018)};
\citet{Salim(2003)};
\citet{Segransan(2000)};
\citet{Shakht(1997)};
\citet{Shkolnik(2012)};
Silverstein, in prep;
\citet{Skrutskie(2006)};
\citet{Smart(2007)};
\citet{Smart(2010b)};
\citet{Soderhjelm(1999)};
\citet{Subasavage(2017)};
\citet{Tanner(2010)};
\citet{Tokovinin(2006b)};
\citet{Tokovinin(2012c)};
\citet{Tokovinin(2016)};
\citet{vanAltena(1995)};
\citet{Vanderspek(2019)};
\citet{vanLeeuwen(2007)};
\citet{Vrijmoet(2020)};
\citet{Wahhaj(2011)};
\citet{Ward-Duong(2015)};
\citet{Weinberger(2016)};
\citet{Weis(1982)};
\citet{Weis(1984)};
\citet{Weis(1986)};
\citet{Weis(1987)};
\citet{Weis(1988)};
\citet{Weis(1991a)};
\citet{Weis(1991b)};
\citet{Weis(1996)};
\citet{Weis(1999)};
\citet{Winters(2011)};
\citet{Winters(2015)};
\citet{Winters(2017)};
\citet{Winters(2018)};
\citet{Winters(2019a)};
\citet{Winters(2019b)};
\citet{Winters(2020a)}.
Winters, in prep.}
\tablecomments{(1) The codes for the techniques and instruments used to detect and resolve systems are: AO det --- adaptive  optics; astdet --- detection via astrometric perturbation,  companion often not detected directly; astorb --- orbit from astrometric measurements; astrad --- orbit from combination of astrometric and radial velocity measurements; eclorb --- eclipsing binary orbit; lkydet --- detection via lucky imaging; radorb --- orbit from radial velocity measurements; spkdet --- detection via speckle interferometry; visdet --- detection via visual astrometry; visorb --- visual orbit.}
\tablecomments{(2) The codes for notes on the DR2 parallax are: w --- weighted mean of two components; A --- adopted parallax and proper motion of primary component; 2p --- 2 parameter astrometric solution; n --- no DR2 data point.}
\tablecomments{(3) 'J' indicates blended photometry.}
\tablecomments{(4) Mass adopted from $^{a}${\citet{Baroch(2018)}};  $^{b}${\citet{Benedict(2001)}};
$^{c}${\citet{Benedict(2016)}}; 
$^{d}${\citet{Duquennoy(1988b)}}; $^{e}${\citet{Eggenberger(2007)}}; $^{f}${\citet{Halbwachs(2000)}}; 
$^{g}${\citet{Han(2002)}}; 
$^{h}${\citet{Heintz(1988)}}; 
$^{i}${\citet{Heintz(1993)}};  
$^{j}${\citet{Heintz(1994)}}; 
$^{k}${\citet{Henry(1993)}}; 
$^{l}${\citet{Henry(1999)}}; $^{m}${\citet{Martinache(2007)}}; $^{n}${\citet{Martinache(2009)}}; $^{o}${\citet{McCarthy(1984)}}; $^{p}${\cite{Morales(2009)}}; $^{q}${\citet{Segransan(2000)}}; $^{r}${\citet{Tokovinin(2006b)}}; 
$^{s}${\citet{Tokovinin(2016)}};  $^{t}${\citet{Winters(2019b)}}; 
$^{u}$Mass calculated from orbital mass ratio; 
$^{v}$Mass from deblended photometry; 
$^{w}$Mass assumed from astrometric orbit.}

\tablecomments{(5) Note on Sample Membership: `s': member of spectroscopic sample; `sc': close companion to member of spectroscopic sample; `mc': close companion to massive primary.}
\tablecomments{This table is available in its entirety in machine-readable form.}
\end{deluxetable*}

\clearpage

\begin{deluxetable*}{clcl}
\tabletypesize{\scriptsize}
\tablecaption{Astrometry, Photometry, and Multiplicity Information for 120 Rejected and Candidate M Dwarfs \label{tab:rejected}}
\tablenum{4}
\tablehead{\colhead{Column}          &
	   \colhead{Format}              &
	   \colhead{Units}               &
	   \colhead{Description}         
	   }
\startdata
1  & A10      & hh:mm:ss         & Right Ascension (J2000.0) \\
2  & A9       & dd:mm:ss         & Declination (J2000.0)    \\
3  & A15      &                  & Name  \\
4  & A1       &                  & Component  \\
5  & A16      &                  & 2MASS Identifier   \\
6  & A5       &                  & Configuration of Multiple Components \\
7  & F10.4    & arcsec           & Separation ($\rho$) Between Multiple Components  \\
8  & A20      &                  & Reference for Multiplicity Detection \\
9  & F4.2     & mag              & Magnitude Difference ($\Delta m$) between Multiple Components  \\
10 & A3       &                  & Filter of Magnitude Difference Multiple Components  \\
11 & A20      &                  & Reference for Multiple Component's  Magnitude Difference \\
12 & F8.4     & mas              & {\it Gaia} DR2 Parallax ($\pi_{DR2}$)  \\
13 & F6.4     & mas              & Uncertainty on Gaia DR2 Parallax \\
14 & A1       &                  & Note on DR2 parallax (1) \\
15 & F6.2     & mas              & Weighted Mean Literature Parallax  ($\pi_{wt}$) \\
16 & F4.2     & mas              & Uncertainty on Literature Parallax \\
17 & A20      &                  & Reference(s) for Parallaxes included in Weighted Mean Parallax   \\
18 & F5.2     & mag              & $G$ Magnitude \\
19 & F4.2     & mag              & $(G_{B_{P}}-G_{R_{P}})$ Color  \\
20 & F5.2     & mag              & $V_{\rm J}$ Magnitude  \\
21 & F5.2     & mag              & $R_{\rm KC}$ Magnitude  \\
22 & F5.2     & mag              & $I_{\rm KC}$ Magnitude   \\
23 & A20      &                  & Reference for $V_{\rm J}R_{\rm KC}I_{\rm KC}$ Magnitudes  \\
24 & F4.1     & mag              & Transformed $R_{KC}$ Magnitude from $G$ and $K_{s}$ \\
25 & F6.3     & mag              & 2MASS $J$ Magnitude   \\
26 & F5.3     & mag              & Uncertainty on 2MASS J Magnitude \\
27 & F6.3     & mag              & 2MASS $H$ Magnitude   \\
28 & F5.3     & mag              & Uncertainty on 2MASS H Magnitude \\
29 & F6.3     & mag              & 2MASS $K_{s}$ Magnitude    \\
30 & F5.3     & mag              & Uncertainty on 2MASS $K_{s}$ Magnitude \\
31 & A1       &                  & Note on 2MASS $K_{s}$ Magnitude (2) \\
32 & F5.3     & M/M$_{\odot}$    & Mass  \\
33 & F5.3     & M/M$_{\odot}$    & Uncertainty on Mass  \\
34 & A1       &                  & Note on Mass (3)  \\
35 & A1       &                  & Note on Sample Membership (4) \\
\enddata
\tablerefs{
This work;          
\citet{Baroch(2018)};
\citet{Bartlett(2017)};
\citet{Benedict(2002c)};
\citet{Bergfors(2010)};
\citet{Bessel(1990)};
\citet{Bessell(1991)};
\citet{Beuzit(2004)};
\citet{Bowler(2015)};
\citet{Burningham(2009)};
\citet{Costa(2005)};
\citet{Dahn(2017)};
\citet{Delfosse(1999c)};
\citet{Dieterich(2012)};
\citet{Dittmann(2014)};
\citet{Doyle(1990)};
\citet{Finch(2016)};
\citet{Finch(2018)};
\citet{Gatewood(2008)};
\citet{Harrington(1985)};
\citet{Heintz(1993)};
\citet{Heintz(1994)};
\citet{Henry(2018)};
\citet{Horch(2015)};
\citet{Hosey(2015)};
\citet{Hummel(1995)};
\citet{Ireland(2008)};
\citet{Janson(2012)};
\citet{Janson(2014a)};
\citet{Jao(2005)};
\citet{Jodar(2013)};
\citet{Khovritchev(2013)};
\citet{Koen(2010)};
\citet{Lindegren(1997)};
\citet{Lindegren(2018)};
\citet{Lurie(2014)};
\citet{Luyten(1997)};
\citet{Reid(2002)};
\citet{Riedel(2010)};
\citet{Riedel(2011)};
\citet{Riedel(2018)};
Silverstein, in prep;              
\citet{Skrutskie(2006)};
\citet{Soderhjelm(1999)};
\citet{vanAltena(1995)};
\citet{vanLeeuwen(2007)};
\citet{Vrijmoet(2020)};
\citet{Weinberger(2016)};
\citet{Weis(1982)};
\citet{Weis(1984)};
\citet{Weis(1986)};
\citet{Weis(1987)};
\citet{Weis(1988)};
\citet{Weis(1991a)};
\citet{Weis(1991b)};
\citet{Weis(1993)};
\citet{Weis(1996)};
\citet{Weis(1999)};
\citet{Winters(2011)};
\citet{Winters(2015)};
\citet{Winters(2017)};
\citet{Winters(2019a)};
\citet{Winters(2020a)};
\citet{Woitas(2003)}.
}
\tablecomments{(1) The codes for notes on the DR2 parallax are: `w' --- weighted mean of two components; `2p' --- 2 parameter astrometric solution.}
\tablecomments{(2) 'J' indicates blended photometry.}
\tablecomments{(3) Mass adopted from $^{a}${\citet{Baroch(2018)}}; $^{b}${\citet{Piccotti(2020)}}; $^{c}$Mass from deblended photometry.}
\tablecomments{(4) Note on Sample Membership: `c': candidate spectroscopic member, `r': rejected.}
\tablecomments{This table is available in its entirety in machine-readable form. }
\end{deluxetable*}

\subsection{Optical \& Infrared Photometry}
\label{subsec:opt_ir_phot}

In order to estimate exposure times for the objects in our  spectroscopic survey, which uses the TiO-bands near 7100~\AA, we required $R_{KC}$ magnitudes. Absolute $V_{\rm J}R_{\rm KC}I_{\rm KC}$\footnote{The central wavelengths for the $V_J$, R$_{KC}$, and I$_{KC}$ filters at the 0.9m are 5438\AA, 6425\AA, and 8075\AA, respectively. We henceforth omit the subscripts on these filters' designations.} photometry for many of the M dwarfs presented here was taken from the literature. The extensive compilations of absolute $VRI$ photometry from Weis \citep{Weis(1982),Weis(1984),Weis(1986),Weis(1987),Weis(1988),Weis(1991a),Weis(1991b),Weis(1993),Weis(1996),Weis(1999)}, Bessel \citep{Bessel(1990),Bessell(1991)}, and SAAO \citep{Koen(2002),Koen(2010),Reid(2002),Reid(2003),Reid(2004)} are valuable resources of M dwarf photometry and provide $VRI$ for 165 of the \coreno ~nearby M dwarfs in our spectroscopic sample presented here. All photometry from Weis has been converted to the Johnson-Kron-Cousins (JKC) system using the relation in \citet{Bessell(1987)}. $VRI$ for an additional 148 objects has been presented previously by RECONS \citep[e.g., in ][]{Costa(2005),Davison(2015),Dieterich(2014),Henry(2004),Henry(2006),Henry(2018),Hosey(2015),Jao(2005),Jao(2011),Jao(2017),Lurie(2014),Mamajek(2013),Riedel(2010),Riedel(2011),Riedel(2014),Vanderspek(2019),Winters(2011),Winters(2015),Winters(2017),Winters(2019a)}.

As part of the long-term effort by the RECONS group at the CTIO/SMARTS 0.9m telescope, we have measured new absolute $VRI$ photometry for 17 objects (plus 3 objects at distances $>$ 15 pc), which is presented here for the first time. Following the methodology outlined in previous RECONS work, various standard star fields from \citet{Graham(1982),Bessel(1990),Landolt(1992),Landolt(2007),Landolt(2013)} were observed multiple times each night to derive transformation equations and extinction curves. Apertures 14\arcsec ~in diameter, matching those used by Landolt, were used to determine the stellar fluxes, except in cases where close contaminating sources required deblending. In these cases, smaller apertures were used and aperture corrections were applied. More details about the data reduction procedures, transformation equations, etc., can be found in \citet{Jao(2005)} and \citet{Winters(2011),Winters(2015)}.


For the remaining 82 stars in our spectroscopic sample with both a $G$ and $R_{KC}$ magnitude\footnote{LHS~1901AB is the only object in the sample that lacks both.}, we created a transformation to estimate an $R_{KC}$ magnitude using a ($G-K_{s}$) color. To create the transformation, we first excluded from the \coreno ~stars in our spectroscopic sample any multiple systems with less than 4\arcsec ~separations, as well as GJ~1194A, which has a blended $R$, but not $K$ magnitude. We also then excluded objects that lack a {\it Gaia} DR2 parallax. While there exists a $G$ magnitude for many of them, these objects are generally binaries with separations less than 1\arcsec. We thus exclude them all under the suspicion that they are all multiple systems. There were three outliers (GJ~1230B, GJ~896B, and GJ~754.1B) that we also excluded from the fit. The final number of M dwarfs used in our fit is 253. We present our second-order fit below:

\begin{equation}
R_{\rm KC} = K_{\rm s} - 0.3098 + 0.8436*(G-K_{\rm s}) + 0.0725*(G-K_{\rm s})^2 
\end{equation}

\noindent We explored higher order fits, but the calculated root-mean-square (RMS) deviation was similar to that of the second order fit presented here; therefore, we adopt the relation with the smallest order. We find a RMS deviation of 0.05 mag. This transformation is valid for M dwarfs with $3.2 < (G-K_{s}) < 5.1$ mag and $2.3 < (B_P-R_P) < 4.5$ mag. We list all photometry, which includes that gathered from the literature, the newly measured magnitudes presented here, and the transformed $R_{KC}$ magnitudes, in Tables \ref{tab:sample} and \ref{tab:rejected}.

\begin{figure}
\hspace{-1.0cm}
\includegraphics[scale=.38,angle=270]{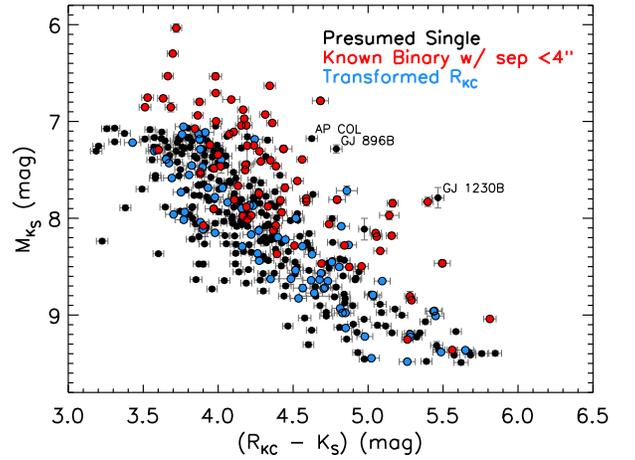}
\caption{An observational Hertzsprung-Russell Diagram for the \coreno ~stars in our spectroscopic sample, except for GJ~1194A and LHS~1901AB. Stars presumed to be single are shown in black, binaries with blended photometry (generally with separations less than 4\arcsec) are noted in red, and objects for which only a transformed $R_{KC}$ is available are shown in blue. Three black points among the elevated red points, two of which are unresolved binary candidates, are labeled. \label{fig:hrd}}
\end{figure}

Figure \ref{fig:hrd} shows our spectroscopic sample plotted on an observational Hertzsprung-Russell Diagram. We have assumed $R-$band uncertainties of 0.05 mag for the transformed magnitudes and 0.03 mag for all others. GJ~1194A is excluded from this plot because, as noted above, its $R$ magnitude is blended, while its $K$ magnitude is resolved. We also excluded LHS~1901AB because it has no published $R_{KC}$ magnitude and no {\it Gaia G} magnitude from which to estimate it. The red points indicate the elevated sequence of known binaries with separations less than 4\arcsec ~from their companions. Three obvious black points among the elevated binaries are AP~COL, GJ~896B, and GJ~1230B. AP~COL is a known nearby young star \citep{Riedel(2011)} \footnote{The few young M dwarfs in this sample, including AP~COL, will have slightly revised masses from the values we present once they are on the main sequence, but they are almost certainly within the mass range of the stars in this paper.}, whereas GJ~896B and GJ~1230B could be as yet unknown unresolved binaries.   

\subsection{Multiples} \label{subsec:mult}

For the stars in Tables \ref{tab:sample} and \ref{tab:rejected}, we provide the multiplicity information for those that are known to host stellar or brown dwarf companions. For the systems that are common to both the objects presented here and those in \citet{Winters(2019a)}, we duplicate the results presented there. The remaining noted multiples are the result of a literature search, which includes previous results from our spectroscopic survey \citep[i.e., from][]{Winters(2018),Winters(2020a)}. 

We note that where orbit determinations from the literature are reported, the semi-major axis, $a$ or $a\sin i$, is listed instead of angular separation $\rho$. If $a$ was not reported in the reference given, it was calculated from the period and the estimated masses of the components in question via Kepler's third law. Magnitude differences from photographic plates are denoted by “V*.” In a few cases, the position angles and/or $\Delta m$ measurements are not available. In many cases, there are multiple separation and $\Delta m$ measurements available in the literature from different groups using different techniques. An exhaustive list of these results is beyond the scope of this work; instead, a single recent result for each system is listed. 

We are systematically searching many of the stars listed in Table \ref{tab:sample} for stellar companions. In our spectroscopic sample, we are surveying all \coreno ~M dwarfs via both high-resolution, multi-epoch spectroscopy and high-resolution speckle imaging. Some of the stars outside our spectroscopic sample have been observed with either or both techniques, as well. It is not our intention to report here the results of these companion searches, but we do note those that we have detected to host a companion. These objects are indicated with a separation of 1\arcsec ~to indicate that a companion has been detected, but not yet resolved, with a reference of Winters et al., in prep. These systems will be described more thoroughly in a future paper.


To date, 74 M dwarf primary stars within 15 pc with masses $0.1 \leq M/\mdot \leq 0.3$ ~are known to host stellar companions at all separations. We can calculate a preliminary and uncorrected stellar multiplicity rate (MR) for the primary stars in our spectroscopic sample, where the MR is the percentage of all systems that are multiple, regardless of whether the system is a binary, triple, or higher-order multiple. We find that $20\pm2$\% ($74/369$) of systems within 15 pc with these types of M dwarf primaries consist of more than one star. Because the MR is known to be a decreasing function of primary mass \citep{Duchene(2013)}, we expect the MR for our mass range of interest to be smaller than that for the entire M dwarf sequence. Comparison to M dwarf MRs reported by others indicates agreement: $26\pm3$\%, $23.5\pm3.2$\%, $26.8\pm1.4$\% \citep{Duchene(2013),Ward-Duong(2015),Winters(2019a)}. Further, the stellar MR of $21.4\pm2.0$\% for the mass range $0.15 - 0.30$ M/\mdot ~reported in \citet{Winters(2019a)}, which is not corrected for incompleteness, is in excellent agreement with the preliminary MR we find here for our target stars of similar mass range ($0.1 \leq M/\mdot \leq 0.3$). When including the five systems within 15 pc known to be composed of an M dwarf primary within our mass range of interest and exclusively brown dwarf (or candidate brown dwarf) companions (GJ~1001, LHS~1610, G~123-45, GJ~595, GJ~1215), we find a preliminary MR of $21\pm2$\% ($79/369$). 

We also find that $43\pm2$\% (222/512) of all M dwarfs with masses $0.1 \leq M/\mdot \leq 0.3$ within 15 pc reside in multiple systems with stars of all masses. That is, while 20\% are hosts to less massive stars, a nearly equal percentage are companions to more massive stars. We emphasize that these MRs are preliminary. We have not completed our high-resolution surveys of these stars, so more companions may be detected in the future.


\subsection{{\it Gaia DR2} Astrometry Comparison} 
\label{subsec:astr_gaia}

\begin{figure}
\hspace{-1.75cm}
\includegraphics[scale=.40,angle=270]{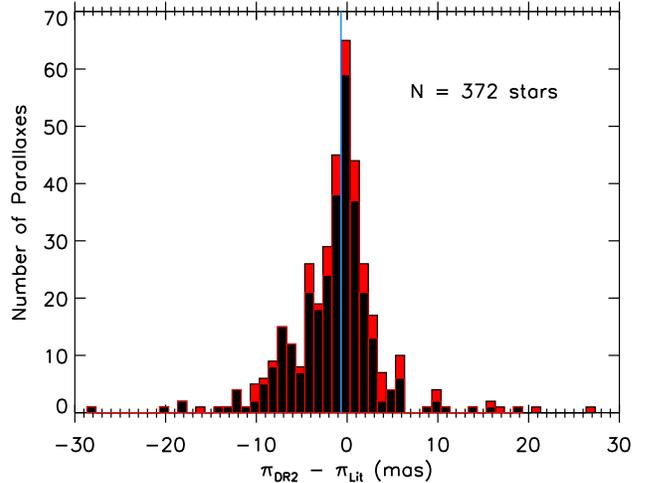}
\caption{Stacked histogram of the parallax differences between {\it Gaia} DR2 results and the weighted mean of parallaxes for the same 372 targets published prior to DR2 in black (noted as `Lit' for literature). The blue vertical line indicates a median offset at -0.67 mas, meaning that DR2 parallaxes are generally smaller (with corresponding distances that are further away) than the parallaxes for the same objects published before the release of DR2 data. In red are noted binaries with separations $<$ 2\arcsec. \label{fig:pi_diff}}
\end{figure}

It is interesting to compare the weighted mean parallaxes available pre-DR2 to the DR2 parallaxes. Of the \coreno ~M dwarfs in our spectroscopic sample, 372 have both a {\it Gaia} DR2 parallax and a parallax that was published before DR2. We show in Figure \ref{fig:pi_diff} the distribution of the difference between the parallax measurements ($\pi_{DR2} - \pi_{wt}$) for these 372 M dwarfs within 15 pc. We note that the distribution is not Gaussian and shows a shoulder in the negative half of the histogram. Because our sample size is relatively small, we found that fitting a curve to the distribution and determining the peak leads to offsets that were dependent on the bin size. We therefore report the median offset, $-0.67^{+0.99}_{-1.79}$ mas. The uncertainty spans $\pm 34$\% of the parallax differences on either side of the median offset ($-2.46$ mas to $0.32$ mas) and is asymmetric due to the asymmetric distribution. This negative difference is in the sense that the DR2 parallaxes are slightly smaller than parallaxes for the same objects available before the DR2, which corresponds to distances slightly further away. This offset is similar to the $-0.71\pm0.06$ mas difference that \citet{Arenou(2018)} found in their comparison of DR2 results to RECONS nearby star parallaxes. The RECONS group conducted their own comparison of {\it Gaia} DR1 parallaxes to $Hipparcos$ parallaxes for stars within 25 pc \citep{Jao(2016)} and find a distance-dependent offset of $-0.24\pm0.02$ mas in their analysis; an offset of similar magnitude is also present in the DR2 data (Jao et al., in prep). Another study compared the distances of eclipsing binaries at 30 pc $-$ 3 kpc with $G <$ 12 mag to their distances derived from DR2 parallaxes and find an offset of $-82\pm33~ \mu$as \citep{Stassun(2018a)}. However, the results of that study are not relevant here, as the stars presented in this work are at closer distances and are thus more sensitive to the distance-dependent offset discussed in \citet{Jao(2016)}. We do not apply any correction to the DR2 parallaxes listed here. The $-0.67$ mas difference we find will not substantially affect any results based on the parallax, as it amounts to 0.3\%  and 1\% of the parallax at 5 and 15 pc, respectively. Future data releases from {\it Gaia} will include longer baselines for the parallax determinations, which will improve the parallax values. Outliers in Figure \ref{fig:pi_diff} at $\pm20-30$ mas are LP~119-26, G~184-31, LTT~13861, and GJ~791.2AB. While GJ~791.2AB and LTT~13861AB are known binaries with a subarcsecond separations between their components, the other two could be unknown binaries.



\begin{figure}
\hspace{-1.75cm}
\includegraphics[scale=.40,angle=270]{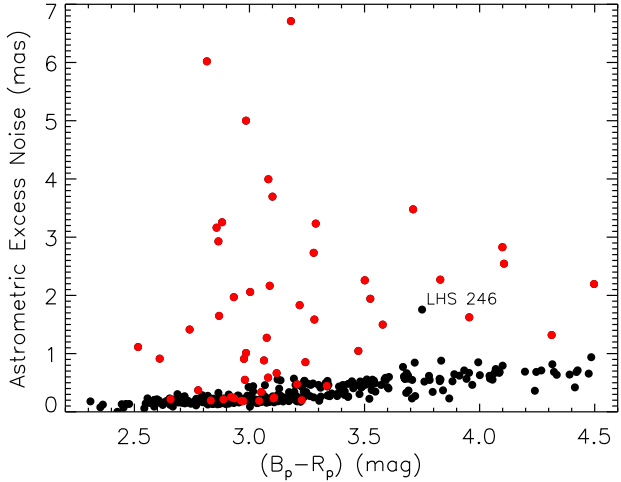}
\caption{{\it Gaia} DR2 astrometric excess noise (aen) as a function of {\it Gaia} ($B_{p}-R_{p}$) color for the 379 M dwarfs with 5-parameter DR2 astrometric solutions.  We indicate binaries with separations $<$ 2\arcsec ~in red, excluding 9 close binaries with weighted mean DR2 parallaxes. The astrometric excess noise increases for redder objects; many close binaries exhibit large astrometric excess noise, due to their treatment as single objects in the DR2 reduction. We label one subarcsecond-separation binary candidate.  \label{fig:ast_noise}}
\end{figure}

As part of the DR2 catalog, astrometric quality flags are provided. For our spectroscopic sample, we assess here two of these flags: the astrometric excess noise (aen) and its significance. As discussed in \citet{Lindegren(2018)} and \citet{Arenou(2018)}, the astrometric excess noise is the goodness of fit or the extra noise, in mas, that is needed to explain the residuals in the astrometric solution for an object. We show in Figure \ref{fig:ast_noise} the DR2 astrometric excess noise as a function of the {\it Gaia} ($B_{p}-R_{p})$ color.  Illustrated is the increase in aen to roughly 1 mas with redder presumably single stars. Binaries with angular separations $<$2\arcsec ~are indicated; many are elevated above the black points which symbolize presumably single stars. One presumably single star lies among the known unresolved binaries and is possibly an unresolved binary itself: LHS~246. 

\begin{figure}
\hspace{-1.75cm}
\includegraphics[scale=.40,angle=270]{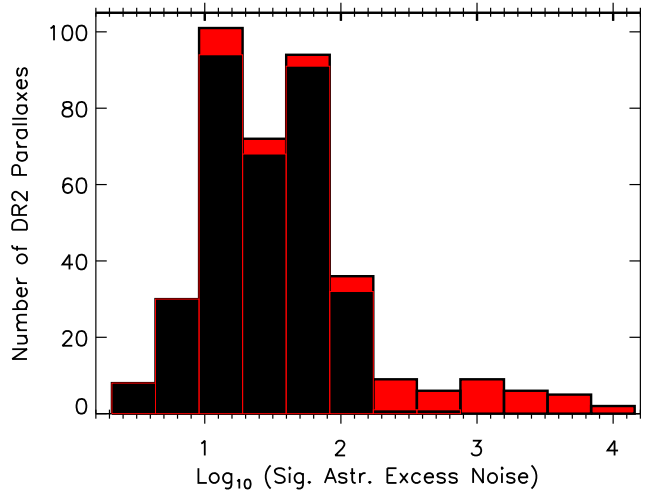}
\caption{Stacked histogram of the significance of the {\it Gaia} DR2 astrometric excess noise for 379 M dwarfs (in black), in logarithmic form. Most of these nearby M dwarfs have astrometric excess noise with significance roughly $10-100$. We show the 56 binaries with separations $<$ 2\arcsec ~that have 5-parameter DR2 astrometric solutions in solid red, not including 9 close binaries with weighted mean DR2 parallaxes. All the binaries with separations $<$ 2\arcsec ~have an astrometric excess noise significance greater than 10, not an unexpected finding.  \label{fig:ast_noise_sig}}
\end{figure}

We also note that for M dwarfs in our selected mass range, the astrometric excess noise significance values are generally large (10-100), as seen in Figure \ref{fig:ast_noise_sig}. While it is noted that the aen significance should be less than 2.0 \citep{Lindegren(2018)}, most of these targets have robust distances via ground-based parallaxes that agree with the DR2 parallaxes. We note binaries with separations $<$2\arcsec. These all have significance values larger than 10, not unexpectedly, as all were treated as single objects in the DR2 astrometric reduction. Presumably, both the astrometric excess noise for these objects and its significance will be reduced in future {\it Gaia} data releases.


\section{Discussion and Conclusion} \label{sec:disc}

We have compiled a volume-complete list of M dwarf primary stars, as well as a list of all the known M dwarf companions, with masses $0.1 \leq M/\mdot \leq 0.3$ ~that lie within 15 pc. This is only now possible by augmenting the parallaxes from the 
{\it Gaia} DR2 and with those from ground-based astrometry efforts. We now investigate the volume-completeness of our sample and whether Eddington or Malmquist biases are present in the sample.

We first compare the actual number of M dwarf primaries within 15 pc within our mass range of interest to the expected number. As mentioned previously, we extrapolate from the number of these types of M dwarf primary stars within 10 pc, assuming completeness within 10 pc and a constant stellar density to 15 pc. As noted by \citet{Henry(2018)}, the 10 pc sample was considered to be at least 90\% complete before the release of parallaxes from the DR2, and the DR2 did not reveal any new M dwarf primaries within this mass range within 10 pc. We therefore assume that the 10 pc sample is complete. There are 104 M dwarf primaries within this mass range and within 10 pc presented here as part of our spectroscopic sample, but there were 109 M dwarfs in the same mass range presented in \citet{Winters(2019a)}. There are two reasons for the slight difference between the two numbers. The first is the availability of DR2 parallaxes, which resulted in some distance shifts for targets near the 10 pc horizon, as well as some mass estimate changes, which affected the membership of some stars. The second reason is that mass estimates differ slightly, depending whether the estimate is from the $M_V$ magnitude \citep[as used in][]{Winters(2019a)} or $M_K$ magnitude, although they agree within their uncertainties. This change in the actual number from 109 to 104 M dwarf primaries within our mass range of interest decreases the expected number within 15 pc from $368\pm19$ to $351\pm19$. The 369 primaries in our sample is in agreement with the expected 351 M dwarf primaries of this kind within 15 pc.



We next assess the completeness of both the M dwarf primary population and the total number of M dwarfs with masses $0.1 \leq M/\mdot ~\leq 0.3$ within the 15 pc volume. In Figures \ref{fig:cum_hist} and \ref{fig:mass_hist}, these groups are noted in blue and green. We do this by comparing the stellar number density for stars of these masses at the 15 pc volume horizon to that of the 10 pc volume. For the 369 M dwarf primaries, the stellar number density at 15 pc is $0.026\pm0.001$ stars per cubic pc, compared to the density of $0.025\pm0.002$ stars per cubic pc for the 104 primaries within 10 pc. These densities agree well. In a similar fashion for the total number (\totalno) of these kinds of M dwarfs within 15 pc, we calculate a stellar number density of $0.036\pm0.002$ stars per cubic parsec, as compared to $0.037\pm0.003$ stars per cubic parsec for the 153 within the 10 pc volume. As with the primaries' density comparison, the densities for the total number of these kinds of M dwarfs within 10 pc and 15 pc are in agreement.

We now investigate whether Eddington and/or Malmquist biases affect various groups of stars that we have presented. The Eddington bias is an over-representation of fainter/more distant objects in samples because, due to random observational errors, their magnitude/parallax uncertainties are larger than those of brighter/nearer stars \citep{Eddington(1913)}. The classical Malmquist bias affects magnitude-limited samples and results in the over-representation of bright stars in samples \citep{Malmquist(1922)}. These biases could potentially affect our sample in both distance and mass space. 

We first investigate whether the Eddington bias affects the stars in our sample in distance space, that is, whether more distant stars make up a disproportionate part of our sample. We perform this analysis by comparing the stellar density of the primary M dwarfs with masses $0.1 \leq M/\mdot \leq 0.3$ ~found within a 1-$\sigma_{\pi}$ volume shell outside of 15 pc to that within the 15-pc volume, assuming a constant stellar density. Within 15 pc, there is one primary star per 38.3 pc$^{3}$. Using the median parallax uncertainty of 0.0805 mas for the primary M dwarfs in our sample, we calculate a volume of 51.4 pc$^{3}$ between 15 pc and 15.017 pc. We thus calculate that there are 1.3 primary stars within the 1-$\sigma$ volume shell outside of 15 pc. We perform a similar analysis for the total number of stars within our mass range of interest and the \coreno ~stars in our spectroscopic sample, where the median parallax uncertainties are 0.0851 mas and 0.0805 mas, respectively. We calculate that 1.9 and 1.5 stars, respectively, lie within the 1-$\sigma_{\pi}$ volume shell outside of 15 pc for those two groups of stars. We therefore conclude that the Eddington bias is not relevant for these three samples (primaries, total, and spectroscopic) in distance space due to the exquisite precision of the DR2 parallaxes.

We next address whether an Eddington bias is present in our sample in mass space, that is, whether the faint, lowest mass primary M dwarfs in our sample are over-represented. As noted above, the masses for most of these M dwarfs are estimated using the MLR in the $M_{K}$ magnitude. Thus, the mass uncertainties are a combination of the uncertainties in the 2MASS $K_{s}$ magnitude, the parallax, and the scatter in the MLR. The median uncertainty in the $K_{s}$ magnitudes for the primaries in our sample is 0.021 mag, and as noted above, the median parallax uncertainty for the primary stars is 0.0805 mas. Because these photometric and astrometric uncertainties are typically small (0.2\% for the faintest target with $K = 10.66\pm0.022$ mag, and 0.12\% for the most distant target at 15 pc), the mass uncertainties are dominated by the 0.014 M/\mdot ~scatter in the MLR. We therefore compare the primary M dwarfs with masses between $0.100-0.114$ M/\mdot, a range of $1-\sigma_{mass}$, within 15 pc to the number of primary stars within this mass range within 10 pc. Within 10 pc are 10 such primaries; extrapolating to 15 pc and assuming a constant stellar density leads us to expect $34\pm6$ such objects. The 39 primary M dwarfs in our sample with masses between $0.100-0.114$ M/\mdot ~are comfortably within the expected value, when considering the Poisson uncertainty. We thus conclude that the lowest mass stars in our sample are not over-represented.

We also perform a similar analysis to determine whether the highest mass primary stars are under- or over-represented in our sample. Because there are only 2 primary stars with masses $0.286-0.300$ M/\mdot ($1-\sigma_{mass}$) within 10 pc, we broaden our comparison range to $2-\sigma_{mass}$ ($0.272-0.300$ M/\mdot) for more reliable statistics. There are 11 primaries in this mass range within 10 pc, from which we extrapolate an expected $37\pm6$ primary stars. The number of primaries in our sample within this mass range within 15 pc (32) is within the expected range, illustrating that neither an Eddington nor a Malmquist bias is present.

The sample of primary stars with masses $0.1 \leq M/\mdot \leq 0.3$ ~within 15 pc that we present here is essentially volume-complete. While there are likely a few additional secondary components to be found within this mass range, we would overall be missing only a few percent of the population of these M dwarfs. Some of these companions we will identify during our high-resolution spectroscopic and speckle imaging surveys. Ultimately, future data releases from Gaia will identify any primary stars within this mass range that we may have missed. These would likely be the primaries of multiple systems with companions at subarcsecond separations that have neither a DR2 parallax nor a previous parallax from the literature. We thus conclude that, to the best of our knowledge, the sample we present here is volume-complete.

These stars are attractive for both stellar astrophysics and exoplanetary science. Thus, the existence of a vetted census of nearby, fully-convective M dwarfs is a valuable resource for those studying them in their own right and those searching them for planets. 

A better understanding of the multiplicity statistics (separation, orbital period, eccentricity, and mass ratio distributions) for this sample of fully-convective stars is valuable for constraining star formation and evolution scenarios. Determining whether a log-normal or a power-law model best fits the orbital period distribution would constrain whether low-mass star formation has a preferred spatial scale or follows a scale-free process in the formation of binaries \citep{Duchene(2013)}. We have identified a number of candidate multiple stars in our spectroscopic sample, which we list in Table \ref{tab:candidates}. Results from our complementary high-resolution spectroscopic and speckle imaging surveys, in combination with {\it Gaia} imaging data and results from previous multiplicity work, will provide a complete picture of the types and separations of companions that are hosted by these fully-convective stars. Moreover, the combination of precise astrometry data from {\it Gaia} with the RV results from our spectroscopic survey will permit a probe of any multiplicity-kinematics/age correlation that could constrain evolution models for low-mass stars. A volume-complete sample eliminates the need for any corrections due to incompleteness.

\begin{deluxetable}{lccc}
\centering
\tabletypesize{\small}
\tablecaption{Candidate Multiple Systems \label{tab:candidates}}
\tablecolumns{4}
\tablenum{5}
\tablehead{\colhead{Name}                &
 	   \colhead{RA}                  &
	   \colhead{DEC}                 &
       \colhead{Flag}            \\
	   \colhead{   }                 &
	   \colhead{(hh:mm:ss)}          &
       \colhead{(dd:mm:ss)}          &
	   \colhead{     }               }
\startdata                
LP~119-26       & 05:16:53.5 & $+$56:40:19   &  $\Delta \pi$ \\  
LEP~J0617+8353  & 06:17:05.3 & $+$83:53:35   &  2p \\
LHS~246         & 08:25:52.8 & $+$69:02:01   &  aen  \\  
LP~549-6        & 09:58:56.5 & $+$05:58:00   &  2p \\
LHS~296         & 11:01:19.7 & $+$03:00:17   &  2p \\
GJ~1227         & 18:22:27.1 & $+$62:03:02   &  2p \\
GJ~1230B        & 18:41:09.8 & $+$24:47:20   &  elev \\  
G~184-31        & 18:49:54.5 & $+$18:40:29   & $\Delta \pi$ \\  
LHS~3746        & 22:02:29.4 & $-$37:04:51   &  2p \\
GJ~896B         & 23:31:52.6 & $+$19:56:14   &  elev \\
\enddata

{\tablecomments{Flag Description: `2p' indicates that only a 2-parameter DR2 astrometric solution is available for the object; `aen' indicates that the DR2 astrometric excess noise for this object is similar to that of known subarcsecond binaries, as illustrated in Figure \ref{fig:ast_noise}; `$\Delta \pi$' indicates that the difference between the object's DR2 and pre-DR2 parallaxes is larger or smaller than $\pm 20$ mas, respectively, as shown in Figure \ref{fig:pi_diff}; `elev' means that the object is elevated above the main sequence in the HR Diagram in Figure \ref{fig:hrd} due to overluminosity.}}
\end{deluxetable}

One of the motivations for planet searches around nearby low-mass stars is to find nearby transiting planets whose atmospheres are most accessible for study. Due to the efforts of MEarth \citep{Nutzman(2008),Irwin(2015)} and $TESS$ \citep{Ricker(2015)}, there are currently six systems among the \coreno ~nearby, M dwarfs presented here that are known to host transiting planets: GJ~1214b \citep{Charbonneau(2009)}, GJ~1132bc \citep{Berta-Thompson(2015), Bonfils(2018)}, LHS~1140bc \citep{Dittmann(2017a), Ment(2019)}, LTT~1445Ab \citep{Winters(2019a)}, LHS~3844b \citep{Vanderspek(2019)}, 2MA~0505-4756b \citep[TOI-540b;][]{Ment(2020)}. Along with the planets orbiting the 0.08 M/\mdot ~ultra-cool dwarf TRAPPIST-1 \citep{Gillon(2017)}, these are some of the most promising small exoplanets for atmospheric studies in the near future with large telescopes.

\vspace{5mm}
\begin{center}
\large
Acknowledgments
\end{center}
\normalsize


We are grateful to Matthew Payne for valuable conversations and suggestions that helped improve the analysis and interpretation of the results presented here. We also thank Eliot Haley Vrijmoet for early access to his excellent M dwarf astrometry paper. We thank the referee for their insightful comments and suggestions that improved the manuscript. We thank Emily Pass for help with refining the sample.

This work is made possible by a grant from the John Templeton Foundation. The opinions expressed in this publication are those of the authors and do not necessarily reflect the views of the John Templeton Foundation. This material is based upon work supported by the National Science Foundation under grant AST-1616624, and work supported by the National Aeronautics and Space Administration under Grant No. 80NSSC18K0476 issued through the XRP Program. The RECONS team is indebted to long-term support from the NSF, most recently under grant AST-1715551. 

We thank the members of the SMARTS Consortium, who have enabled the operations of the small telescopes at CTIO since 2003, as well as observers and observer support at CTIO, specifically Arturo Gomez, Mauricio Rojas, Hernan Tirado, Joselino Vasquez, Alberto Miranda, and Edgardo Cosgrove. 

Data products from the Two Micron All Sky Survey, which is a joint
project of the University of Massachusetts and the Infrared Processing
and Analysis Center/California Institute of Technology, funded by the
National Aeronautics and Space Administration (NASA) and the NSF have
been used extensively, as have the SIMBAD database and the Aladin and
Vizier interfaces, operated at CDS, Strasbourg, France. This work has
made ample use of the Smithsonian Astrophysical Observatory/NASA
Astrophysics Data System. This work has made use of data from the European Space Agency (ESA) mission {\it Gaia} (\url{https://www.cosmos.esa.int/gaia}), processed by the {\it Gaia} Data Processing and Analysis Consortium (DPAC, \url{https://www.cosmos.esa.int/web/gaia/dpac/consortium}). Funding for the DPAC has been provided by national institutions, in particular the institutions participating in the {\it Gaia} Multilateral Agreement. This research has made use of the Washington Double Star Catalog maintained at the U.S. Naval Observatory.

\facilities{CTIO:0.9m}

\software{IRAF \citep{Tody(1986),Tody(1993)}, IDL, python}

\clearpage


\bibliographystyle{aasjournal}
\bibliography{masterref.bib}

\begin{thebibliography}{}
\expandafter\ifx\csname natexlab\endcsname\relax\def\natexlab#1{#1}\fi
\providecommand{\url}[1]{\href{#1}{#1}}

\bibitem[{{Allen} {et~al.}(2012){Allen}, {Burgasser}, {Faherty}, \&
  {Kirkpatrick}}]{Allen(2012)}
{Allen}, P.~R., {Burgasser}, A.~J., {Faherty}, J.~K., \& {Kirkpatrick}, J.~D.
  2012, \aj, 144, 62

\bibitem[{{Arenou} {et~al.}(2018){Arenou}, {Luri}, {Babusiaux}, {Fabricius},
  {Helmi}, {Muraveva}, {Robin}, {Spoto}, {Vallenari}, {Antoja},
  {Cantat-Gaudin}, {Jordi}, {Leclerc}, {Reyl{\'e}}, {Romero-G{\'o}mez}, {Shih},
  {Soria}, {Barache}, {Bossini}, {Bragaglia}, {Breddels}, {Fabrizio},
  {Lambert}, {Marrese}, {Massari}, {Moitinho}, {Robichon}, {Ruiz-Dern},
  {Sordo}, {Veljanoski}, {Eyer}, {Jasniewicz}, {Pancino}, {Soubiran}, {Spagna},
  {Tanga}, {Turon}, \& {Zurbach}}]{Arenou(2018)}
{Arenou}, F., {Luri}, X., {Babusiaux}, C., {et~al.} 2018, \aap, 616, A17

\bibitem[{{Balega} {et~al.}(2013){Balega}, {Balega}, {Gasanova}, {Dyachenko},
  {Maksimov}, {Malogolovets}, {Rastegaev}, \& {Shkhagosheva}}]{Balega(2013)}
{Balega}, I.~I., {Balega}, Y.~Y., {Gasanova}, L.~T., {et~al.} 2013,
  Astrophysical Bulletin, 68, 53

\bibitem[{{Baroch} {et~al.}(2018){Baroch}, {Morales}, {Ribas}, {Tal-Or},
  {Zechmeister}, {Reiners}, {Caballero}, {Quirrenbach}, {Amado}, {Dreizler},
  {Lalitha}, {Jeffers}, {Lafarga}, {B{\'e}jar}, {Colom{\'e}},
  {Cort{\'e}s-Contreras}, {D{\'\i}ez-Alonso}, {Galad{\'\i}-Enr{\'\i}quez},
  {Guenther}, {Hagen}, {Henning}, {Herrero}, {K{\"u}rster}, {Montes}, {Nagel},
  {Passegger}, {Perger}, {Rosich}, {Schweitzer}, \& {Seifert}}]{Baroch(2018)}
{Baroch}, D., {Morales}, J.~C., {Ribas}, I., {et~al.} 2018, \aap, 619, A32

\bibitem[{{Bartlett} {et~al.}(2017){Bartlett}, {Lurie}, {Riedel}, {Ianna},
  {Jao}, {Henry}, {Winters}, {Finch}, \& {Subasavage}}]{Bartlett(2017)}
{Bartlett}, J.~L., {Lurie}, J.~C., {Riedel}, A., {et~al.} 2017, \aj, 154, 151

\bibitem[{{Benedict} {et~al.}(2000){Benedict}, {McArthur}, {Franz},
  {Wasserman}, \& {Henry}}]{Benedict(2000b)}
{Benedict}, G.~F., {McArthur}, B.~E., {Franz}, O.~G., {Wasserman}, L.~H., \&
  {Henry}, T.~J. 2000, \aj, 120, 1106

\bibitem[{{Benedict} {et~al.}(1999){Benedict}, {McArthur}, {Chappell}, {Nelan},
  {Jefferys}, {van Altena}, {Lee}, {Cornell}, {Shelus}, {Hemenway}, {Franz},
  {Wasserman}, {Duncombe}, {Story}, {Whipple}, \& {Fredrick}}]{Benedict(1999)}
{Benedict}, G.~F., {McArthur}, B., {Chappell}, D.~W., {et~al.} 1999, \aj, 118,
  1086

\bibitem[{{Benedict} {et~al.}(2001){Benedict}, {McArthur}, {Franz},
  {Wasserman}, {Henry}, {Takato}, {Strateva}, {Crawford}, {Ianna}, {McCarthy},
  {Nelan}, {Jefferys}, {van Altena}, {Shelus}, {Hemenway}, {Duncombe}, {Story},
  {Whipple}, {Bradley}, \& {Fredrick}}]{Benedict(2001)}
{Benedict}, G.~F., {McArthur}, B.~E., {Franz}, O.~G., {et~al.} 2001, \aj, 121,
  1607

\bibitem[{{Benedict} {et~al.}(2002){Benedict}, {McArthur}, {Forveille},
  {Delfosse}, {Nelan}, {Butler}, {Spiesman}, {Marcy}, {Goldman}, {Perrier},
  {Jefferys}, \& {Mayor}}]{Benedict(2002c)}
{Benedict}, G.~F., {McArthur}, B.~E., {Forveille}, T., {et~al.} 2002, \apjl,
  581, L115

\bibitem[{{Benedict} {et~al.}(2016){Benedict}, {Henry}, {Franz}, {McArthur},
  {Wasserman}, {Jao}, {Cargile}, {Dieterich}, {Bradley}, {Nelan}, \&
  {Whipple}}]{Benedict(2016)}
{Benedict}, G.~F., {Henry}, T.~J., {Franz}, O.~G., {et~al.} 2016, \aj, 152, 141

\bibitem[{{Bergfors} {et~al.}(2010){Bergfors}, {Brandner}, {Janson}, {Daemgen},
  {Geissler}, {Henning}, {Hippler}, {Hormuth}, {Joergens}, \&
  {K{\"o}hler}}]{Bergfors(2010)}
{Bergfors}, C., {Brandner}, W., {Janson}, M., {et~al.} 2010, \aap, 520, A54

\bibitem[{{Berta-Thompson} {et~al.}(2015){Berta-Thompson}, {Irwin},
  {Charbonneau}, {Newton}, {Dittmann}, {Astudillo-Defru}, {Bonfils}, {Gillon},
  {Jehin}, {Stark}, {Stalder}, {Bouchy}, {Delfosse}, {Forveille}, {Lovis},
  {Mayor}, {Neves}, {Pepe}, {Santos}, {Udry}, \&
  {W{\"u}nsche}}]{Berta-Thompson(2015)}
{Berta-Thompson}, Z.~K., {Irwin}, J., {Charbonneau}, D., {et~al.} 2015, \nat,
  527, 204

\bibitem[{{Bessel}(1990)}]{Bessel(1990)}
{Bessel}, M.~S. 1990, \aaps, 83, 357

\bibitem[{{Bessell}(1991)}]{Bessell(1991)}
{Bessell}, M.~S. 1991, \aj, 101, 662

\bibitem[{{Bessell} \& {Weis}(1987)}]{Bessell(1987)}
{Bessell}, M.~S., \& {Weis}, E.~W. 1987, \pasp, 99, 642

\bibitem[{{Beuzit} {et~al.}(2004){Beuzit}, {S{\'e}gransan}, {Forveille},
  {Udry}, {Delfosse}, {Mayor}, {Perrier}, {Hainaut}, {Roddier}, {Roddier}, \&
  {Mart{\'{\i}}n}}]{Beuzit(2004)}
{Beuzit}, J.-L., {S{\'e}gransan}, D., {Forveille}, T., {et~al.} 2004, \aap,
  425, 997

\bibitem[{{Bochanski} {et~al.}(2010){Bochanski}, {Hawley}, {Covey}, {West},
  {Reid}, {Golimowski}, \& {Ivezi{\'c}}}]{Bochanski(2010)}
{Bochanski}, J.~J., {Hawley}, S.~L., {Covey}, K.~R., {et~al.} 2010, \aj, 139,
  2679

\bibitem[{{Bonfils} {et~al.}(2018){Bonfils}, {Almenara}, {Cloutier},
  {W{\"u}nsche}, {Astudillo-Defru}, {Berta-Thompson}, {Bouchy}, {Charbonneau},
  {Delfosse}, \& {D{\'\i}az}}]{Bonfils(2018)}
{Bonfils}, X., {Almenara}, J.~M., {Cloutier}, R., {et~al.} 2018, \aap, 618,
  A142

\bibitem[{{Bowler} {et~al.}(2015){Bowler}, {Liu}, {Shkolnik}, \&
  {Tamura}}]{Bowler(2015)}
{Bowler}, B.~P., {Liu}, M.~C., {Shkolnik}, E.~L., \& {Tamura}, M. 2015, \apjs,
  216, 7

\bibitem[{{Burningham} {et~al.}(2009){Burningham}, {Pinfield}, {Leggett},
  {Tinney}, {Liu}, {Homeier}, {West}, {Day-Jones}, {Huelamo}, {Dupuy}, {Zhang},
  {Murray}, {Lodieu}, {Barrado Y Navascu{\'e}s}, {Folkes}, {Galvez-Ortiz},
  {Jones}, {Lucas}, {Calderon}, \& {Tamura}}]{Burningham(2009)}
{Burningham}, B., {Pinfield}, D.~J., {Leggett}, S.~K., {et~al.} 2009, \mnras,
  395, 1237

\bibitem[{{Chabrier}(2003)}]{Chabrier(2003b)}
{Chabrier}, G. 2003, \pasp, 115, 763

\bibitem[{{Charbonneau} \& {Deming}(2007)}]{Charbonneau(2007)}
{Charbonneau}, D., \& {Deming}, D. 2007, arXiv e-prints, arXiv:0706.1047

\bibitem[{{Charbonneau} {et~al.}(2009){Charbonneau}, {Berta}, {Irwin}, {Burke},
  {Nutzman}, {Buchhave}, {Lovis}, {Bonfils}, {Latham}, {Udry}, {Murray-Clay},
  {Holman}, {Falco}, {Winn}, {Queloz}, {Pepe}, {Mayor}, {Delfosse}, \&
  {Forveille}}]{Charbonneau(2009)}
{Charbonneau}, D., {Berta}, Z.~K., {Irwin}, J., {et~al.} 2009, \nat, 462, 891

\bibitem[{{Cort{\'e}s-Contreras} {et~al.}(2017){Cort{\'e}s-Contreras},
  {B{\'e}jar}, {Caballero}, {Gauza}, {Montes}, {Alonso-Floriano}, {Jeffers},
  {Morales}, {Reiners}, {Ribas}, {Sch{\"o}fer}, {Quirrenbach}, {Amado},
  {Mundt}, \& {Seifert}}]{Cortes-Contreras(2017)}
{Cort{\'e}s-Contreras}, M., {B{\'e}jar}, V.~J.~S., {Caballero}, J.~A., {et~al.}
  2017, \aap, 597, A47

\bibitem[{{Costa} {et~al.}(2005){Costa}, {M{\'e}ndez}, {Jao}, {Henry},
  {Subasavage}, {Brown}, {Ianna}, \& {Bartlett}}]{Costa(2005)}
{Costa}, E., {M{\'e}ndez}, R.~A., {Jao}, W.-C., {et~al.} 2005, \aj, 130, 337

\bibitem[{{Cutri} {et~al.}(2003){Cutri}, {Skrutskie}, {van Dyk}, {Beichman},
  {Carpenter}, {Chester}, {Cambresy}, {Evans}, {Fowler}, {Gizis}, {Howard},
  {Huchra}, {Jarrett}, {Kopan}, {Kirkpatrick}, {Light}, {Marsh}, {McCallon},
  {Schneider}, {Stiening}, {Sykes}, {Weinberg}, {Wheaton}, {Wheelock}, \&
  {Zacarias}}]{Cutri(2003)}
{Cutri}, R.~M., {Skrutskie}, M.~F., {van Dyk}, S., {et~al.} 2003, VizieR Online
  Data Catalog, II/246

\bibitem[{{Dahn} {et~al.}(1988){Dahn}, {Harrington}, {Kallarakal}, {Guetter},
  {Luginbuhl}, {Riepe}, {Walker}, {Pier}, {Vrba}, {Monet}, \&
  {Ables}}]{Dahn(1988)}
{Dahn}, C.~C., {Harrington}, R.~S., {Kallarakal}, V.~V., {et~al.} 1988, \aj,
  95, 237

\bibitem[{{Dahn} {et~al.}(2017){Dahn}, {Harris}, {Subasavage}, {Ables},
  {Canzian}, {Guetter}, {Harris}, {Henden}, {Leggett}, {Levine}, {Luginbuhl},
  {Monet}, {Monet}, {Munn}, {Pier}, {Stone}, {Vrba}, {Walker}, \&
  {Tilleman}}]{Dahn(2017)}
{Dahn}, C.~C., {Harris}, H.~C., {Subasavage}, J.~P., {et~al.} 2017, \aj, 154,
  147

\bibitem[{{Davison} {et~al.}(2015){Davison}, {White}, {Henry}, {Riedel}, {Jao},
  {Bailey}, {Quinn}, {Cantrell}, {Subasavage}, \& {Winters}}]{Davison(2015)}
{Davison}, C.~L., {White}, R.~J., {Henry}, T.~J., {et~al.} 2015, \aj, 149, 106

\bibitem[{{Delfosse} {et~al.}(1999{\natexlab{a}}){Delfosse}, {Forveille},
  {Beuzit}, {Udry}, {Mayor}, \& {Perrier}}]{Delfosse(1999c)}
{Delfosse}, X., {Forveille}, T., {Beuzit}, J.-L., {et~al.} 1999{\natexlab{a}},
  \aap, 344, 897

\bibitem[{{Delfosse} {et~al.}(2000){Delfosse}, {Forveille}, {S{\'e}gransan},
  {Beuzit}, {Udry}, {Perrier}, \& {Mayor}}]{Delfosse(2000)}
{Delfosse}, X., {Forveille}, T., {S{\'e}gransan}, D., {et~al.} 2000, \aap, 364,
  217

\bibitem[{{Delfosse} {et~al.}(1999{\natexlab{b}}){Delfosse}, {Forveille},
  {Udry}, {Beuzit}, {Mayor}, \& {Perrier}}]{Delfosse(1999d)}
{Delfosse}, X., {Forveille}, T., {Udry}, S., {et~al.} 1999{\natexlab{b}}, \aap,
  350, L39

\bibitem[{{Dieterich} {et~al.}(2012){Dieterich}, {Henry}, {Golimowski},
  {Krist}, \& {Tanner}}]{Dieterich(2012)}
{Dieterich}, S.~B., {Henry}, T.~J., {Golimowski}, D.~A., {Krist}, J.~E., \&
  {Tanner}, A.~M. 2012, \aj, 144, 64

\bibitem[{{Dieterich} {et~al.}(2014){Dieterich}, {Henry}, {Jao}, {Winters},
  {Hosey}, {Riedel}, \& {Subasavage}}]{Dieterich(2014)}
{Dieterich}, S.~B., {Henry}, T.~J., {Jao}, W.-C., {et~al.} 2014, \aj, 147, 94

\bibitem[{{Dittmann} {et~al.}(2014){Dittmann}, {Irwin}, {Charbonneau}, \&
  {Berta-Thompson}}]{Dittmann(2014)}
{Dittmann}, J.~A., {Irwin}, J.~M., {Charbonneau}, D., \& {Berta-Thompson},
  Z.~K. 2014, \apj, 784, 156

\bibitem[{{Dittmann} {et~al.}(2017){Dittmann}, {Irwin}, {Charbonneau},
  {Bonfils}, {Astudillo-Defru}, {Haywood}, {Berta-Thompson}, {Newton},
  {Rodriguez}, {Winters}, {Tan}, {Almenara}, {Bouchy}, {Delfosse}, {Forveille},
  {Lovis}, {Murgas}, {Pepe}, {Santos}, {Udry}, {W{\"u}nsche}, {Esquerdo},
  {Latham}, \& {Dressing}}]{Dittmann(2017a)}
{Dittmann}, J.~A., {Irwin}, J.~M., {Charbonneau}, D., {et~al.} 2017, \nat, 544,
  333

\bibitem[{{Docobo} {et~al.}(2006){Docobo}, {Tamazian}, {Balega}, \&
  {Melikian}}]{Docobo(2006a)}
{Docobo}, J.~A., {Tamazian}, V.~S., {Balega}, Y.~Y., \& {Melikian}, N.~D. 2006,
  \aj, 132, 994

\bibitem[{{Doyle} \& {Butler}(1990)}]{Doyle(1990)}
{Doyle}, J.~G., \& {Butler}, C.~J. 1990, \aap, 235, 335

\bibitem[{{Duch{\^e}ne} \& {Kraus}(2013)}]{Duchene(2013)}
{Duch{\^e}ne}, G., \& {Kraus}, A. 2013, \araa, 51, 269

\bibitem[{{Dupuy} \& {Liu}(2017)}]{Dupuy(2017)}
{Dupuy}, T.~J., \& {Liu}, M.~C. 2017, \apjs, 231, 15

\bibitem[{{Duquennoy} \& {Mayor}(1988)}]{Duquennoy(1988b)}
{Duquennoy}, A., \& {Mayor}, M. 1988, \aap, 200, 135

\bibitem[{{Eddington}(1913)}]{Eddington(1913)}
{Eddington}, A.~S. 1913, \mnras, 73, 359

\bibitem[{{Eggenberger} {et~al.}(2007){Eggenberger}, {Udry}, {Chauvin},
  {Beuzit}, {Lagrange}, {S{\'e}gransan}, \& {Mayor}}]{Eggenberger(2007)}
{Eggenberger}, A., {Udry}, S., {Chauvin}, G., {et~al.} 2007, \aap, 474, 273

\bibitem[{{Fabricius} \& {Makarov}(2000)}]{Fabricius(2000)}
{Fabricius}, C., \& {Makarov}, V.~V. 2000, \aaps, 144, 45

\bibitem[{{Faherty} {et~al.}(2012){Faherty}, {Burgasser}, {Walter}, {Van der
  Bliek}, {Shara}, {Cruz}, {West}, {Vrba}, \&
  {Anglada-Escud{\'e}}}]{Faherty(2012)}
{Faherty}, J.~K., {Burgasser}, A.~J., {Walter}, F.~M., {et~al.} 2012, \apj,
  752, 56

\bibitem[{{Farihi} {et~al.}(2005){Farihi}, {Becklin}, \&
  {Zuckerman}}]{Farihi(2005)}
{Farihi}, J., {Becklin}, E.~E., \& {Zuckerman}, B. 2005, \apjs, 161, 394

\bibitem[{{Finch} \& {Zacharias}(2016)}]{Finch(2016)}
{Finch}, C.~T., \& {Zacharias}, N. 2016, \aj, 151, 160

\bibitem[{{Finch} {et~al.}(2018){Finch}, {Zacharias}, \& {Jao}}]{Finch(2018)}
{Finch}, C.~T., {Zacharias}, N., \& {Jao}, W.-C. 2018, \aj, 155, 176

\bibitem[{{Gaia Collaboration} {et~al.}(2016){Gaia Collaboration}, {Prusti},
  {de Bruijne}, {Brown}, {Vallenari}, {Babusiaux}, {Bailer-Jones}, {Bastian},
  {Biermann}, {Evans}, \& et~al.}]{Gaia(2016a)}
{Gaia Collaboration}, {Prusti}, T., {de Bruijne}, J.~H.~J., {et~al.} 2016,
  \aap, 595, A1

\bibitem[{{Gaia Collaboration} {et~al.}(2018){Gaia Collaboration}, {Brown},
  {Vallenari}, {Prusti}, {de Bruijne}, {Babusiaux}, {Bailer-Jones}, {Biermann},
  {Evans}, {Eyer}, \& et~al.}]{GaiaDR2(2018)}
{Gaia Collaboration}, {Brown}, A.~G.~A., {Vallenari}, A., {et~al.} 2018, \aap,
  616, A1

\bibitem[{{Gatewood}(2008)}]{Gatewood(2008)}
{Gatewood}, G. 2008, \aj, 136, 452

\bibitem[{{Gatewood} \& {Coban}(2009)}]{Gatewood(2009)}
{Gatewood}, G., \& {Coban}, L. 2009, \aj, 137, 402

\bibitem[{{Gatewood} {et~al.}(2003){Gatewood}, {Coban}, \&
  {Han}}]{Gatewood(2003)}
{Gatewood}, G., {Coban}, L., \& {Han}, I. 2003, \aj, 125, 1530

\bibitem[{{Gillon} {et~al.}(2017){Gillon}, {Demory}, {Van Grootel}, {Motalebi},
  {Lovis}, {Cameron}, {Charbonneau}, {Latham}, {Molinari}, {Pepe},
  {S{\'e}gransan}, {Sasselov}, {Udry}, {Mayor}, {Micela}, {Piotto}, \&
  {Sozzetti}}]{Gillon(2017)}
{Gillon}, M., {Demory}, B.-O., {Van Grootel}, V., {et~al.} 2017, Nature
  Astronomy, 1, 0056

\bibitem[{{Golimowski} {et~al.}(2004){Golimowski}, {Leggett}, {Marley}, {Fan},
  {Geballe}, {Knapp}, {Vrba}, {Henden}, {Luginbuhl}, {Guetter}, {Munn},
  {Canzian}, {Zheng}, {Tsvetanov}, {Chiu}, {Glazebrook}, {Hoversten},
  {Schneider}, \& {Brinkmann}}]{Golimowski(2004)}
{Golimowski}, D.~A., {Leggett}, S.~K., {Marley}, M.~S., {et~al.} 2004, \aj,
  127, 3516

\bibitem[{{Graham}(1982)}]{Graham(1982)}
{Graham}, J.~A. 1982, \pasp, 94, 244

\bibitem[{{Halbwachs} {et~al.}(2000){Halbwachs}, {Arenou}, {Mayor}, {Udry}, \&
  {Queloz}}]{Halbwachs(2000)}
{Halbwachs}, J.~L., {Arenou}, F., {Mayor}, M., {Udry}, S., \& {Queloz}, D.
  2000, \aap, 355, 581

\bibitem[{{Halley Vrijmoet} {et~al.}(2020){Halley Vrijmoet}, {Henry}, {Jao}, \&
  {Dieterich}}]{Vrijmoet(2020)}
{Halley Vrijmoet}, E., {Henry}, T.~J., {Jao}, W.-C., \& {Dieterich}, S.~B.
  2020, arXiv e-prints, arXiv:2009.00121

\bibitem[{{Han} \& {Gatewood}(2002)}]{Han(2002)}
{Han}, I., \& {Gatewood}, G. 2002, \pasp, 114, 224

\bibitem[{{Harrington} {et~al.}(1981){Harrington}, {Christy}, \&
  {Strand}}]{Harrington(1981)}
{Harrington}, R.~S., {Christy}, J.~W., \& {Strand}, K.~A. 1981, \aj, 86, 909

\bibitem[{{Harrington} {et~al.}(1985){Harrington}, {Kallarakal}, {Christy},
  {Dahn}, {Riepe}, {Guetter}, {Ables}, {Hewitt}, {Vrba}, \&
  {Walker}}]{Harrington(1985)}
{Harrington}, R.~S., {Kallarakal}, V.~V., {Christy}, J.~W., {et~al.} 1985, \aj,
  90, 123

\bibitem[{{Heintz}(1974)}]{Heintz(1974)}
{Heintz}, W.~D. 1974, \aj, 79, 819

\bibitem[{{Heintz}(1988)}]{Heintz(1988)}
---. 1988, \aaps, 72, 543

\bibitem[{{Heintz}(1990)}]{Heintz(1990)}
---. 1990, \apjs, 74, 275

\bibitem[{{Heintz}(1991)}]{Heintz(1991)}
---. 1991, \aj, 101, 1071

\bibitem[{{Heintz}(1993)}]{Heintz(1993)}
---. 1993, \aj, 105, 1188

\bibitem[{{Heintz}(1994)}]{Heintz(1994)}
---. 1994, \aj, 108, 2338

\bibitem[{{Henry} {et~al.}(1999){Henry}, {Franz}, {Wasserman}, {Benedict},
  {Shelus}, {Ianna}, {Kirkpatrick}, \& {McCarthy}}]{Henry(1999)}
{Henry}, T.~J., {Franz}, O.~G., {Wasserman}, L.~H., {et~al.} 1999, \apj, 512,
  864

\bibitem[{{Henry} {et~al.}(1997){Henry}, {Ianna}, {Kirkpatrick}, \&
  {Jahreiss}}]{Henry(1997)}
{Henry}, T.~J., {Ianna}, P.~A., {Kirkpatrick}, J.~D., \& {Jahreiss}, H. 1997,
  \aj, 114, 388

\bibitem[{{Henry} {et~al.}(2006){Henry}, {Jao}, {Subasavage}, {Beaulieu},
  {Ianna}, {Costa}, \& {M{\'e}ndez}}]{Henry(2006)}
{Henry}, T.~J., {Jao}, W.-C., {Subasavage}, J.~P., {et~al.} 2006, \aj, 132,
  2360

\bibitem[{{Henry} \& {McCarthy}(1993)}]{Henry(1993)}
{Henry}, T.~J., \& {McCarthy}, Jr., D.~W. 1993, \aj, 106, 773

\bibitem[{{Henry} {et~al.}(2004){Henry}, {Subasavage}, {Brown}, {Beaulieu},
  {Jao}, \& {Hambly}}]{Henry(2004)}
{Henry}, T.~J., {Subasavage}, J.~P., {Brown}, M.~A., {et~al.} 2004, \aj, 128,
  2460

\bibitem[{{Henry} {et~al.}(2018){Henry}, {Jao}, {Winters}, {Dieterich},
  {Finch}, {Ianna}, {Riedel}, {Silverstein}, {Subasavage}, \&
  {Vrijmoet}}]{Henry(2018)}
{Henry}, T.~J., {Jao}, W.-C., {Winters}, J.~G., {et~al.} 2018, \aj, 155, 265

\bibitem[{{Hershey} \& {Taff}(1998)}]{Hershey(1998)}
{Hershey}, J.~L., \& {Taff}, L.~G. 1998, \aj, 116, 1440

\bibitem[{{H{\o}g} {et~al.}(2000){H{\o}g}, {Fabricius}, {Makarov}, {Urban},
  {Corbin}, {Wycoff}, {Bastian}, {Schwekendiek}, \& {Wicenec}}]{Hog(2000)}
{H{\o}g}, E., {Fabricius}, C., {Makarov}, V.~V., {et~al.} 2000, \aap, 355, L27

\bibitem[{{Horch} {et~al.}(2012){Horch}, {Bahi}, {Gaulin}, {Howell}, {Sherry},
  {Baena Gall{\'e}}, \& {van Altena}}]{Horch(2012a)}
{Horch}, E.~P., {Bahi}, L.~A.~P., {Gaulin}, J.~R., {et~al.} 2012, \aj, 143, 10

\bibitem[{{Horch} {et~al.}(2010){Horch}, {Falta}, {Anderson}, {DeSousa},
  {Miniter}, {Ahmed}, \& {van Altena}}]{Horch(2010)}
{Horch}, E.~P., {Falta}, D., {Anderson}, L.~M., {et~al.} 2010, \aj, 139, 205

\bibitem[{{Horch} {et~al.}(2011){Horch}, {Gomez}, {Sherry}, {Howell}, {Ciardi},
  {Anderson}, \& {van Altena}}]{Horch(2011a)}
{Horch}, E.~P., {Gomez}, S.~C., {Sherry}, W.~H., {et~al.} 2011, \aj, 141, 45

\bibitem[{{Horch} {et~al.}(2015){Horch}, {van Altena}, {Demarque}, {Howell},
  {Everett}, {Ciardi}, {Teske}, {Henry}, \& {Winters}}]{Horch(2015)}
{Horch}, E.~P., {van Altena}, W.~F., {Demarque}, P., {et~al.} 2015, \aj, 149,
  151

\bibitem[{{Hosey} {et~al.}(2015){Hosey}, {Henry}, {Jao}, {Dieterich},
  {Winters}, {Lurie}, {Riedel}, \& {Subasavage}}]{Hosey(2015)}
{Hosey}, A.~D., {Henry}, T.~J., {Jao}, W.-C., {et~al.} 2015, \aj, 150, 6

\bibitem[{{Hummel} {et~al.}(1995){Hummel}, {Armstrong}, {Buscher},
  {Mozurkewich}, {Quirrenbach}, \& {Vivekanand}}]{Hummel(1995)}
{Hummel}, C.~A., {Armstrong}, J.~T., {Buscher}, D.~F., {et~al.} 1995, \aj, 110,
  376

\bibitem[{{Ianna} {et~al.}(1996){Ianna}, {Patterson}, \& {Swain}}]{Ianna(1996)}
{Ianna}, P.~A., {Patterson}, R.~J., \& {Swain}, M.~A. 1996, \aj, 111, 492

\bibitem[{{Ireland} {et~al.}(2008){Ireland}, {Kraus}, {Martinache}, {Lloyd}, \&
  {Tuthill}}]{Ireland(2008)}
{Ireland}, M.~J., {Kraus}, A., {Martinache}, F., {Lloyd}, J.~P., \& {Tuthill},
  P.~G. 2008, \apj, 678, 463

\bibitem[{{Irwin} {et~al.}(2015){Irwin}, {Berta-Thompson}, {Charbonneau},
  {Dittmann}, {Falco}, {Newton}, \& {Nutzman}}]{Irwin(2015)}
{Irwin}, J.~M., {Berta-Thompson}, Z.~K., {Charbonneau}, D., {et~al.} 2015, in
  Cambridge Workshop on Cool Stars, Stellar Systems, and the Sun, Vol.~18, 18th
  Cambridge Workshop on Cool Stars, Stellar Systems, and the Sun, ed. G.~T.
  {van Belle} \& H.~C. {Harris}, 767--772

\bibitem[{{Janson} {et~al.}(2014){Janson}, {Bergfors}, {Brandner},
  {Kudryavtseva}, {Hormuth}, {Hippler}, \& {Henning}}]{Janson(2014a)}
{Janson}, M., {Bergfors}, C., {Brandner}, W., {et~al.} 2014, \apj, 789, 102

\bibitem[{{Janson} {et~al.}(2012){Janson}, {Hormuth}, {Bergfors}, {Brandner},
  {Hippler}, {Daemgen}, {Kudryavtseva}, {Schmalzl}, {Schnupp}, \&
  {Henning}}]{Janson(2012)}
{Janson}, M., {Hormuth}, F., {Bergfors}, C., {et~al.} 2012, \apj, 754, 44

\bibitem[{{Jao} {et~al.}(2016){Jao}, {Henry}, {Riedel}, {Winters}, {Slatten},
  \& {Gies}}]{Jao(2016)}
{Jao}, W.-C., {Henry}, T.~J., {Riedel}, A.~R., {et~al.} 2016, \apjl, 832, L18

\bibitem[{{Jao} {et~al.}(2003){Jao}, {Henry}, {Subasavage}, {Bean}, {Costa},
  {Ianna}, \& {M{\'e}ndez}}]{Jao(2003)}
{Jao}, W.-C., {Henry}, T.~J., {Subasavage}, J.~P., {et~al.} 2003, \aj, 125, 332

\bibitem[{{Jao} {et~al.}(2005){Jao}, {Henry}, {Subasavage}, {Brown}, {Ianna},
  {Bartlett}, {Costa}, \& {M{\'e}ndez}}]{Jao(2005)}
---. 2005, \aj, 129, 1954

\bibitem[{{Jao} {et~al.}(2011){Jao}, {Henry}, {Subasavage}, {Winters},
  {Riedel}, \& {Ianna}}]{Jao(2011)}
---. 2011, \aj, 141, 117

\bibitem[{{Jao} {et~al.}(2017){Jao}, {Henry}, {Winters}, {Subasavage},
  {Riedel}, {Silverstein}, \& {Ianna}}]{Jao(2017)}
{Jao}, W.-C., {Henry}, T.~J., {Winters}, J.~G., {et~al.} 2017, \aj, 154, 191

\bibitem[{{J{\'o}dar} {et~al.}(2013){J{\'o}dar}, {P{\'e}rez-Garrido},
  {D{\'{\i}}az-S{\'a}nchez}, {Vill{\'o}}, {Rebolo}, \&
  {P{\'e}rez-Prieto}}]{Jodar(2013)}
{J{\'o}dar}, E., {P{\'e}rez-Garrido}, A., {D{\'{\i}}az-S{\'a}nchez}, A.,
  {et~al.} 2013, \mnras, 429, 859

\bibitem[{{Khovritchev} {et~al.}(2013){Khovritchev}, {Izmailov}, \&
  {Khrutskaya}}]{Khovritchev(2013)}
{Khovritchev}, M.~Y., {Izmailov}, I.~S., \& {Khrutskaya}, E.~V. 2013, \mnras,
  435, 1083

\bibitem[{{Koen} {et~al.}(2002){Koen}, {Kilkenny}, {van Wyk}, {Cooper}, \&
  {Marang}}]{Koen(2002)}
{Koen}, C., {Kilkenny}, D., {van Wyk}, F., {Cooper}, D., \& {Marang}, F. 2002,
  \mnras, 334, 20

\bibitem[{{Koen} {et~al.}(2010){Koen}, {Kilkenny}, {van Wyk}, \&
  {Marang}}]{Koen(2010)}
{Koen}, C., {Kilkenny}, D., {van Wyk}, F., \& {Marang}, F. 2010, \mnras, 403,
  1949

\bibitem[{{K{\"o}hler} {et~al.}(2012){K{\"o}hler}, {Ratzka}, \&
  {Leinert}}]{Kohler(2012)}
{K{\"o}hler}, R., {Ratzka}, T., \& {Leinert}, C. 2012, \aap, 541, A29

\bibitem[{{Lacy}(1977)}]{Lacy(1977)}
{Lacy}, C.~H. 1977, \apj, 218, 444

\bibitem[{{Landolt}(1992)}]{Landolt(1992)}
{Landolt}, A.~U. 1992, \aj, 104, 372

\bibitem[{{Landolt}(2007)}]{Landolt(2007)}
---. 2007, \aj, 133, 2502

\bibitem[{{Landolt}(2013)}]{Landolt(2013)}
---. 2013, \aj, 146, 131

\bibitem[{{Law} {et~al.}(2008){Law}, {Hodgkin}, \& {Mackay}}]{Law(2008)}
{Law}, N.~M., {Hodgkin}, S.~T., \& {Mackay}, C.~D. 2008, \mnras, 384, 150

\bibitem[{{Leinert} {et~al.}(1994){Leinert}, {Weitzel}, {Richichi}, {Eckart},
  \& {Tacconi-Garman}}]{Leinert(1994)}
{Leinert}, C., {Weitzel}, N., {Richichi}, A., {Eckart}, A., \&
  {Tacconi-Garman}, L.~E. 1994, \aap, 291, L47

\bibitem[{{L{\`e}pine} \& {Shara}(2005)}]{Lepine(2005a)}
{L{\`e}pine}, S., \& {Shara}, M.~M. 2005, \aj, 129, 1483

\bibitem[{{L{\`e}pine} {et~al.}(2009){L{\`e}pine}, {Thorstensen}, {Shara}, \&
  {Rich}}]{Lepine(2009)}
{L{\`e}pine}, S., {Thorstensen}, J.~R., {Shara}, M.~M., \& {Rich}, R.~M. 2009,
  \aj, 137, 4109

\bibitem[{{Lindegren} {et~al.}(1997){Lindegren}, {Mignard}, {S{\"o}derhjelm},
  {Badiali}, {Bernstein}, {Lampens}, {Pannunzio}, {Arenou}, {Bernacca},
  {Falin}, {Froeschl{\'e}}, {Kovalevsky}, {Martin}, {Perryman}, \&
  {Wielen}}]{Lindegren(1997)}
{Lindegren}, L., {Mignard}, F., {S{\"o}derhjelm}, S., {et~al.} 1997, \aap, 323,
  L53

\bibitem[{{Lindegren} {et~al.}(2018){Lindegren}, {Hern{\'a}ndez}, {Bombrun},
  {Klioner}, {Bastian}, {Ramos-Lerate}, {de Torres}, {Steidelm{\"u}ller},
  {Stephenson}, {Hobbs}, {Lammers}, {Biermann}, {Geyer}, {Hilger}, {Michalik},
  {Stampa}, {McMillan}, {Casta{\~n}eda}, {Clotet}, {Comoretto}, {Davidson},
  {Fabricius}, {Gracia}, {Hambly}, {Hutton}, {Mora}, {Portell}, {van Leeuwen},
  {Abbas}, {Abreu}, {Altmann}, {Andrei}, {Anglada}, {Balaguer-N{\'u}{\~n}ez},
  {Barache}, {Becciani}, {Bertone}, {Bianchi}, {Bouquillon}, {Bourda},
  {Br{\"u}semeister}, {Bucciarelli}, {Busonero}, {Buzzi}, {Cancelliere},
  {Carlucci}, {Charlot}, {Cheek}, {Crosta}, {Crowley}, {de Bruijne}, {de
  Felice}, {Drimmel}, {Esquej}, {Fienga}, {Fraile}, {Gai}, {Garralda},
  {Gonz{\'a}lez-Vidal}, {Guerra}, {Hauser}, {Hofmann}, {Holl}, {Jordan},
  {Lattanzi}, {Lenhardt}, {Liao}, {Licata}, {Lister}, {L{\"o}ffler},
  {Marchant}, {Martin-Fleitas}, {Messineo}, {Mignard}, {Morbidelli}, {Poggio},
  {Riva}, {Rowell}, {Salguero}, {Sarasso}, {Sciacca}, {Siddiqui}, {Smart},
  {Spagna}, {Steele}, {Taris}, {Torra}, {van Elteren}, {van Reeven}, \&
  {Vecchiato}}]{Lindegren(2018)}
{Lindegren}, L., {Hern{\'a}ndez}, J., {Bombrun}, A., {et~al.} 2018, \aap, 616,
  A2

\bibitem[{{L{\'o}pez-Morales} {et~al.}(2019){L{\'o}pez-Morales}, {Ben-Ami},
  {Gonzalez-Abad}, {Garc{\'\i}a-Mej{\'\i}a}, {Dietrich}, \&
  {Szentgyorgyi}}]{Lopez-Morales(2019)}
{L{\'o}pez-Morales}, M., {Ben-Ami}, S., {Gonzalez-Abad}, G., {et~al.} 2019,
  \aj, 158, 24

\bibitem[{{Lowrance} {et~al.}(2002){Lowrance}, {Kirkpatrick}, \&
  {Beichman}}]{Lowrance(2002)}
{Lowrance}, P.~J., {Kirkpatrick}, J.~D., \& {Beichman}, C.~A. 2002, \apjl, 572,
  L79

\bibitem[{{Luri} {et~al.}(2018){Luri}, {Brown}, {Sarro}, {Arenou},
  {Bailer-Jones}, {Castro-Ginard}, {de Bruijne}, {Prusti}, {Babusiaux}, \&
  {Delgado}}]{Luri(2018)}
{Luri}, X., {Brown}, A.~G.~A., {Sarro}, L.~M., {et~al.} 2018, \aap, 616, A9

\bibitem[{{Lurie} {et~al.}(2014){Lurie}, {Henry}, {Jao}, {Quinn}, {Winters},
  {Ianna}, {Koerner}, {Riedel}, \& {Subasavage}}]{Lurie(2014)}
{Lurie}, J.~C., {Henry}, T.~J., {Jao}, W.-C., {et~al.} 2014, \aj, 148, 91

\bibitem[{{Luyten}(1979)}]{Luyten(1979b)}
{Luyten}, W.~J. 1979, {NLTT catalogue. Volume\_I. +90\_\_to\_+30\_.
  Volume.\_II. +30\_\_to\_0\_.} (University of Minnesota)

\bibitem[{{Luyten}(1997)}]{Luyten(1997)}
---. 1997, VizieR Online Data Catalog, I/130

\bibitem[{{Malmquist}(1922)}]{Malmquist(1922)}
{Malmquist}, K.~G. 1922, Meddelanden fran Lunds Astronomiska Observatorium
  Serie I, 100, 1

\bibitem[{{Mamajek} {et~al.}(2013){Mamajek}, {Bartlett}, {Seifahrt}, {Henry},
  {Dieterich}, {Lurie}, {Kenworthy}, {Jao}, {Riedel}, {Subasavage}, {Winters},
  {Finch}, {Ianna}, \& {Bean}}]{Mamajek(2013)}
{Mamajek}, E.~E., {Bartlett}, J.~L., {Seifahrt}, A., {et~al.} 2013, \aj, 146,
  154

\bibitem[{{Martin} \& {Mignard}(1998)}]{Martin(1998a)}
{Martin}, C., \& {Mignard}, F. 1998, \aap, 330, 585

\bibitem[{{Martinache} {et~al.}(2007){Martinache}, {Lloyd}, {Ireland},
  {Yamada}, \& {Tuthill}}]{Martinache(2007)}
{Martinache}, F., {Lloyd}, J.~P., {Ireland}, M.~J., {Yamada}, R.~S., \&
  {Tuthill}, P.~G. 2007, \apj, 661, 496

\bibitem[{{Martinache} {et~al.}(2009){Martinache}, {Rojas-Ayala}, {Ireland},
  {Lloyd}, \& {Tuthill}}]{Martinache(2009)}
{Martinache}, F., {Rojas-Ayala}, B., {Ireland}, M.~J., {Lloyd}, J.~P., \&
  {Tuthill}, P.~G. 2009, \apj, 695, 1183

\bibitem[{{Mason} {et~al.}(2018){Mason}, {Hartkopf}, {Miles}, {Subasavage},
  {Raghavan}, \& {Henry}}]{Mason(2018)}
{Mason}, B.~D., {Hartkopf}, W.~I., {Miles}, K.~N., {et~al.} 2018, \aj, 155, 215

\bibitem[{{Mason} {et~al.}(2019){Mason}, {Wycoff}, {Hartkopf}, {Douglass}, \&
  {Worley}}]{Mason(2019)}
{Mason}, B.~D., {Wycoff}, G.~L., {Hartkopf}, W.~I., {Douglass}, G.~G., \&
  {Worley}, C.~E. 2019, VizieR Online Data Catalog, B/wds

\bibitem[{{McArthur} {et~al.}(2010){McArthur}, {Benedict}, {Barnes},
  {Martioli}, {Korzennik}, {Nelan}, \& {Butler}}]{McArthur(2010)}
{McArthur}, B.~E., {Benedict}, G.~F., {Barnes}, R., {et~al.} 2010, \apj, 715,
  1203

\bibitem[{{McCarthy}(1984)}]{McCarthy(1984)}
{McCarthy}, D.~W., J. 1984, \aj, 89, 433

\bibitem[{{Medina} {et~al.}(2020){Medina}, {Winters}, {Irwin}, \&
  {Charbonneau}}]{Medina(2020)}
{Medina}, A.~A., {Winters}, J.~G., {Irwin}, J.~M., \& {Charbonneau}, D. 2020,
  arXiv e-prints, arXiv:2010.15635

\bibitem[{{Ment} {et~al.}(2019){Ment}, {Dittmann}, {Astudillo-Defru},
  {Charbonneau}, {Irwin}, {Bonfils}, {Murgas}, {Almenara}, {Forveille}, {Agol},
  {Ballard}, {Berta-Thompson}, {Bouchy}, {Cloutier}, {Delfosse}, {Doyon},
  {Dressing}, {Esquerdo}, {Haywood}, {Kipping}, {Latham}, {Lovis}, {Newton},
  {Pepe}, {Rodriguez}, {Santos}, {Tan}, {Udry}, {Winters}, \&
  {W{\"u}nsche}}]{Ment(2019)}
{Ment}, K., {Dittmann}, J.~A., {Astudillo-Defru}, N., {et~al.} 2019, \aj, 157,
  32

\bibitem[{{Ment} {et~al.}(2020){Ment}, {Irwin}, {Charbonneau}, {Winters},
  {Medina}, {Cloutier}, {D{\'\i}az}, {Jenkins}, {Ziegler}, {Law}, {Mann},
  {Ricker}, {Vanderspek}, {Latham}, {Seager}, {Winn}, {Jenkins}, {Goeke},
  {Levine}, {Rojas-Ayala}, {Rowden}, {Ting}, \& {Twicken}}]{Ment(2020)}
{Ment}, K., {Irwin}, J., {Charbonneau}, D., {et~al.} 2020, arXiv e-prints,
  arXiv:2009.13623

\bibitem[{{Montagnier} {et~al.}(2006){Montagnier}, {S{\'e}gransan}, {Beuzit},
  {Forveille}, {Delorme}, {Delfosse}, {Perrier}, {Udry}, {Mayor}, {Chauvin},
  {Lagrange}, {Mouillet}, {Fusco}, {Gigan}, \& {Stadler}}]{Montagnier(2006)}
{Montagnier}, G., {S{\'e}gransan}, D., {Beuzit}, J.-L., {et~al.} 2006, \aap,
  460, L19

\bibitem[{{Morales} {et~al.}(2009){Morales}, {Ribas}, {Jordi}, {Torres},
  {Gallardo}, {Guinan}, {Charbonneau}, {Wolf}, {Latham}, {Anglada-Escud{\'e}},
  {Bradstreet}, {Everett}, {O'Donovan}, {Mandushev}, \&
  {Mathieu}}]{Morales(2009)}
{Morales}, J.~C., {Ribas}, I., {Jordi}, C., {et~al.} 2009, \apj, 691, 1400

\bibitem[{{Morley} {et~al.}(2017){Morley}, {Kreidberg}, {Rustamkulov},
  {Robinson}, \& {Fortney}}]{Morley(2017)}
{Morley}, C.~V., {Kreidberg}, L., {Rustamkulov}, Z., {Robinson}, T., \&
  {Fortney}, J.~J. 2017, \apj, 850, 121

\bibitem[{{Nutzman} \& {Charbonneau}(2008)}]{Nutzman(2008)}
{Nutzman}, P., \& {Charbonneau}, D. 2008, \pasp, 120, 317

\bibitem[{{Perryman} {et~al.}(1997){Perryman}, {Lindegren}, {Kovalevsky},
  {Hoeg}, {Bastian}, {Bernacca}, {Cr{\'e}z{\'e}}, {Donati}, {Grenon},
  {Grewing}, {van Leeuwen}, {van der Marel}, {Mignard}, {Murray}, {Le Poole},
  {Schrijver}, {Turon}, {Arenou}, {Froeschl{\'e}}, \&
  {Petersen}}]{Perryman(1997)}
{Perryman}, M.~A.~C., {Lindegren}, L., {Kovalevsky}, J., {et~al.} 1997, \aap,
  323, L49

\bibitem[{{Piccotti} {et~al.}(2020){Piccotti}, {Docobo}, {Carini}, {Tamazian},
  {Brocato}, {Andrade}, \& {Campo}}]{Piccotti(2020)}
{Piccotti}, L., {Docobo}, J.~{\'A}., {Carini}, R., {et~al.} 2020, \mnras, 492,
  2709

\bibitem[{{Pravdo} {et~al.}(2004){Pravdo}, {Shaklan}, {Henry}, \&
  {Benedict}}]{Pravdo(2004)}
{Pravdo}, S.~H., {Shaklan}, S.~B., {Henry}, T., \& {Benedict}, G.~F. 2004,
  \apj, 617, 1323

\bibitem[{{Reid} {et~al.}(2002){Reid}, {Kilkenny}, \& {Cruz}}]{Reid(2002)}
{Reid}, I.~N., {Kilkenny}, D., \& {Cruz}, K.~L. 2002, \aj, 123, 2822

\bibitem[{{Reid} {et~al.}(2003){Reid}, {Cruz}, {Allen}, {Mungall}, {Kilkenny},
  {Liebert}, {Hawley}, {Fraser}, {Covey}, \& {Lowrance}}]{Reid(2003)}
{Reid}, I.~N., {Cruz}, K.~L., {Allen}, P., {et~al.} 2003, \aj, 126, 3007

\bibitem[{{Reid} {et~al.}(2004){Reid}, {Cruz}, {Allen}, {Mungall}, {Kilkenny},
  {Liebert}, {Hawley}, {Fraser}, {Covey}, {Lowrance}, {Kirkpatrick}, \&
  {Burgasser}}]{Reid(2004)}
---. 2004, \aj, 128, 463

\bibitem[{{Ricker} {et~al.}(2015){Ricker}, {Winn}, {Vanderspek}, {Latham},
  {Bakos}, {Bean}, {Berta-Thompson}, {Brown}, {Buchhave}, {Butler}, {Butler},
  {Chaplin}, {Charbonneau}, {Christensen-Dalsgaard}, {Clampin}, {Deming},
  {Doty}, {De Lee}, {Dressing}, {Dunham}, {Endl}, {Fressin}, {Ge}, {Henning},
  {Holman}, {Howard}, {Ida}, {Jenkins}, {Jernigan}, {Johnson}, {Kaltenegger},
  {Kawai}, {Kjeldsen}, {Laughlin}, {Levine}, {Lin}, {Lissauer}, {MacQueen},
  {Marcy}, {McCullough}, {Morton}, {Narita}, {Paegert}, {Palle}, {Pepe},
  {Pepper}, {Quirrenbach}, {Rinehart}, {Sasselov}, {Sato}, {Seager},
  {Sozzetti}, {Stassun}, {Sullivan}, {Szentgyorgyi}, {Torres}, {Udry}, \&
  {Villasenor}}]{Ricker(2015)}
{Ricker}, G.~R., {Winn}, J.~N., {Vanderspek}, R., {et~al.} 2015, Journal of
  Astronomical Telescopes, Instruments, and Systems, 1, 014003

\bibitem[{{Riedel} {et~al.}(2011){Riedel}, {Murphy}, {Henry}, {Melis}, {Jao},
  \& {Subasavage}}]{Riedel(2011)}
{Riedel}, A.~R., {Murphy}, S.~J., {Henry}, T.~J., {et~al.} 2011, \aj, 142, 104

\bibitem[{{Riedel} {et~al.}(2018){Riedel}, {Silverstein}, {Henry}, {Jao},
  {Winters}, {Subasavage}, {Malo}, \& {Hambly}}]{Riedel(2018)}
{Riedel}, A.~R., {Silverstein}, M.~L., {Henry}, T.~J., {et~al.} 2018, \aj, 156,
  49

\bibitem[{{Riedel} {et~al.}(2010){Riedel}, {Subasavage}, {Finch}, {Jao},
  {Henry}, {Winters}, {Brown}, {Ianna}, {Costa}, \& {Mendez}}]{Riedel(2010)}
{Riedel}, A.~R., {Subasavage}, J.~P., {Finch}, C.~T., {et~al.} 2010, \aj, 140,
  897

\bibitem[{{Riedel} {et~al.}(2014){Riedel}, {Finch}, {Henry}, {Subasavage},
  {Jao}, {Malo}, {Rodriguez}, {White}, {Gies}, {Dieterich}, {Winters},
  {Davison}, {Nelan}, {Blunt}, {Cruz}, {Rice}, \& {Ianna}}]{Riedel(2014)}
{Riedel}, A.~R., {Finch}, C.~T., {Henry}, T.~J., {et~al.} 2014, \aj, 147, 85

\bibitem[{{Rodler} \& {L{\'o}pez-Morales}(2014)}]{Rodler(2014)}
{Rodler}, F., \& {L{\'o}pez-Morales}, M. 2014, \apj, 781, 54

\bibitem[{{Salim} \& {Gould}(2003)}]{Salim(2003)}
{Salim}, S., \& {Gould}, A. 2003, \apj, 582, 1011

\bibitem[{{S{\'e}gransan} {et~al.}(2000){S{\'e}gransan}, {Delfosse},
  {Forveille}, {Beuzit}, {Udry}, {Perrier}, \& {Mayor}}]{Segransan(2000)}
{S{\'e}gransan}, D., {Delfosse}, X., {Forveille}, T., {et~al.} 2000, \aap, 364,
  665

\bibitem[{{Shakht}(1997)}]{Shakht(1997)}
{Shakht}, N.~A. 1997, Astronomical and Astrophysical Transactions, 13, 327

\bibitem[{{Shkolnik} {et~al.}(2012){Shkolnik}, {Anglada-Escud{\'e}}, {Liu},
  {Bowler}, {Weinberger}, {Boss}, {Reid}, \& {Tamura}}]{Shkolnik(2012)}
{Shkolnik}, E.~L., {Anglada-Escud{\'e}}, G., {Liu}, M.~C., {et~al.} 2012, \apj,
  758, 56

\bibitem[{{Skrutskie} {et~al.}(2006){Skrutskie}, {Cutri}, {Stiening},
  {Weinberg}, {Schneider}, {Carpenter}, {Beichman}, {Capps}, {Chester},
  {Elias}, {Huchra}, {Liebert}, {Lonsdale}, {Monet}, {Price}, {Seitzer},
  {Jarrett}, {Kirkpatrick}, {Gizis}, {Howard}, {Evans}, {Fowler}, {Fullmer},
  {Hurt}, {Light}, {Kopan}, {Marsh}, {McCallon}, {Tam}, {Van Dyk}, \&
  {Wheelock}}]{Skrutskie(2006)}
{Skrutskie}, M.~F., {Cutri}, R.~M., {Stiening}, R., {et~al.} 2006, \aj, 131,
  1163

\bibitem[{{Smart} {et~al.}(2010){Smart}, {Ioannidis}, {Jones}, {Bucciarelli},
  \& {Lattanzi}}]{Smart(2010b)}
{Smart}, R.~L., {Ioannidis}, G., {Jones}, H.~R.~A., {Bucciarelli}, B., \&
  {Lattanzi}, M.~G. 2010, \aap, 514, A84

\bibitem[{{Smart} {et~al.}(2007){Smart}, {Lattanzi}, {Jahrei{\ss}},
  {Bucciarelli}, \& {Massone}}]{Smart(2007)}
{Smart}, R.~L., {Lattanzi}, M.~G., {Jahrei{\ss}}, H., {Bucciarelli}, B., \&
  {Massone}, G. 2007, \aap, 464, 787

\bibitem[{{Snellen} {et~al.}(2013){Snellen}, {de Kok}, {le Poole}, {Brogi}, \&
  {Birkby}}]{Snellen(2013)}
{Snellen}, I.~A.~G., {de Kok}, R.~J., {le Poole}, R., {Brogi}, M., \& {Birkby},
  J. 2013, \apj, 764, 182

\bibitem[{{S{\"o}derhjelm}(1999)}]{Soderhjelm(1999)}
{S{\"o}derhjelm}, S. 1999, \aap, 341, 121

\bibitem[{{Stassun} \& {Torres}(2018)}]{Stassun(2018a)}
{Stassun}, K.~G., \& {Torres}, G. 2018, \apj, 862, 61

\bibitem[{{Subasavage} {et~al.}(2017){Subasavage}, {Jao}, {Henry}, {Harris},
  {Dahn}, {Bergeron}, {Dufour}, {Dunlap}, {Barlow}, {Ianna}, {L{\'e}pine}, \&
  {Margheim}}]{Subasavage(2017)}
{Subasavage}, J.~P., {Jao}, W.-C., {Henry}, T.~J., {et~al.} 2017, \aj, 154, 32

\bibitem[{{Tanner} {et~al.}(2010){Tanner}, {Gelino}, \& {Law}}]{Tanner(2010)}
{Tanner}, A.~M., {Gelino}, C.~R., \& {Law}, N.~M. 2010, \pasp, 122, 1195

\bibitem[{{Tody}(1986)}]{Tody(1986)}
{Tody}, D. 1986, in \procspie, Vol. 627, Instrumentation in astronomy VI, ed.
  D.~L. {Crawford}, 733

\bibitem[{{Tody}(1993)}]{Tody(1993)}
{Tody}, D. 1993, in Astronomical Society of the Pacific Conference Series,
  Vol.~52, Astronomical Data Analysis Software and Systems II, ed. R.~J.
  {Hanisch}, R.~J.~V. {Brissenden}, \& J.~{Barnes}, 173

\bibitem[{{Tokovinin} \& {Horch}(2016)}]{Tokovinin(2016)}
{Tokovinin}, A., \& {Horch}, E.~P. 2016, \aj, 152, 116

\bibitem[{{Tokovinin} \& {L{\'e}pine}(2012)}]{Tokovinin(2012c)}
{Tokovinin}, A., \& {L{\'e}pine}, S. 2012, \aj, 144, 102

\bibitem[{{Tokovinin} {et~al.}(2006){Tokovinin}, {Thomas}, {Sterzik}, \&
  {Udry}}]{Tokovinin(2006b)}
{Tokovinin}, A., {Thomas}, S., {Sterzik}, M., \& {Udry}, S. 2006, \aap, 450,
  681

\bibitem[{{van Altena} {et~al.}(1995){van Altena}, {Lee}, \&
  {Hoffleit}}]{vanAltena(1995)}
{van Altena}, W.~F., {Lee}, J.~T., \& {Hoffleit}, D. 1995, VizieR Online Data
  Catalog, 1174, 0

\bibitem[{{van de Kamp}(1971)}]{vandeKamp(1971)}
{van de Kamp}, P. 1971, \araa, 9, 103

\bibitem[{{van Leeuwen}(2007)}]{vanLeeuwen(2007)}
{van Leeuwen}, F. 2007, \aap, 474, 653

\bibitem[{{Vanderspek} {et~al.}(2019){Vanderspek}, {Huang}, {Vanderburg},
  {Ricker}, {Latham}, {Seager}, {Winn}, {Jenkins}, {Burt}, {Dittmann},
  {Newton}, {Quinn}, {Shporer}, {Charbonneau}, {Irwin}, {Ment}, {Winters},
  {Collins}, {Evans}, {Gan}, {Hart}, {Jensen}, {Kielkopf}, {Mao}, {Waalkes},
  {Bouchy}, {Marmier}, {Nielsen}, {Ottoni}, {Pepe}, {S{\'e}gransan}, {Udry},
  {Henry}, {Paredes}, {James}, {Hinojosa}, {Silverstein}, {Palle},
  {Berta-Thompson}, {Crossfield}, {Davies}, {Dragomir}, {Fausnaugh}, {Glidden},
  {Pepper}, {Morgan}, {Rose}, {Twicken}, {Villase{\~n}or}, {Yu}, {Bakos},
  {Bean}, {Buchhave}, {Christensen-Dalsgaard}, {Christiansen}, {Ciardi},
  {Clampin}, {De Lee}, {Deming}, {Doty}, {Jernigan}, {Kaltenegger}, {Lissauer},
  {McCullough}, {Narita}, {Paegert}, {Pal}, {Rinehart}, {Sasselov}, {Sato},
  {Sozzetti}, {Stassun}, \& {Torres}}]{Vanderspek(2019)}
{Vanderspek}, R., {Huang}, C.~X., {Vanderburg}, A., {et~al.} 2019, \apjl, 871,
  L24

\bibitem[{{Wahhaj} {et~al.}(2011){Wahhaj}, {Liu}, {Biller}, {Clarke},
  {Nielsen}, {Close}, {Hayward}, {Mamajek}, {Cushing}, {Dupuy}, {Tecza},
  {Thatte}, {Chun}, {Ftaclas}, {Hartung}, {Reid}, {Shkolnik}, {Alencar},
  {Artymowicz}, {Boss}, {de Gouveia Dal Pino}, {Gregorio-Hetem}, {Ida},
  {Kuchner}, {Lin}, \& {Toomey}}]{Wahhaj(2011)}
{Wahhaj}, Z., {Liu}, M.~C., {Biller}, B.~A., {et~al.} 2011, \apj, 729, 139

\bibitem[{{Ward-Duong} {et~al.}(2015){Ward-Duong}, {Patience}, {De Rosa},
  {Bulger}, {Rajan}, {Goodwin}, {Parker}, {McCarthy}, \&
  {Kulesa}}]{Ward-Duong(2015)}
{Ward-Duong}, K., {Patience}, J., {De Rosa}, R.~J., {et~al.} 2015, \mnras, 449,
  2618

\bibitem[{{Weinberger} {et~al.}(2016){Weinberger}, {Boss}, {Keiser},
  {Anglada-Escud{\'e}}, {Thompson}, \& {Burley}}]{Weinberger(2016)}
{Weinberger}, A.~J., {Boss}, A.~P., {Keiser}, S.~A., {et~al.} 2016, \aj, 152,
  24

\bibitem[{{Weis}(1984)}]{Weis(1984)}
{Weis}, E.~W. 1984, \apjs, 55, 289

\bibitem[{{Weis}(1986)}]{Weis(1986)}
---. 1986, \aj, 91, 626

\bibitem[{{Weis}(1987)}]{Weis(1987)}
---. 1987, \aj, 93, 451

\bibitem[{{Weis}(1988)}]{Weis(1988)}
---. 1988, \apss, 142, 223

\bibitem[{{Weis}(1991{\natexlab{a}})}]{Weis(1991a)}
---. 1991{\natexlab{a}}, \aj, 101, 1882

\bibitem[{{Weis}(1991{\natexlab{b}})}]{Weis(1991b)}
---. 1991{\natexlab{b}}, \aj, 102, 1795

\bibitem[{{Weis}(1993)}]{Weis(1993)}
---. 1993, \aj, 105, 1962

\bibitem[{{Weis}(1996)}]{Weis(1996)}
---. 1996, \aj, 112, 2300

\bibitem[{{Weis}(1999)}]{Weis(1999)}
---. 1999, \aj, 117, 3021

\bibitem[{{Weis} \& {Upgren}(1982)}]{Weis(1982)}
{Weis}, E.~W., \& {Upgren}, A.~R. 1982, \pasp, 94, 821

\bibitem[{{Winters} {et~al.}(2011){Winters}, {Henry}, {Jao}, {Subasavage},
  {Finch}, \& {Hambly}}]{Winters(2011)}
{Winters}, J.~G., {Henry}, T.~J., {Jao}, W.-C., {et~al.} 2011, \aj, 141, 21

\bibitem[{{Winters} {et~al.}(2015){Winters}, {Henry}, {Lurie}, {Hambly}, {Jao},
  {Bartlett}, {Boyd}, {Dieterich}, {Finch}, {Hosey}, {Ianna}, {Riedel},
  {Slatten}, \& {Subasavage}}]{Winters(2015)}
{Winters}, J.~G., {Henry}, T.~J., {Lurie}, J.~C., {et~al.} 2015, \aj, 149, 5

\bibitem[{{Winters} {et~al.}(2017){Winters}, {Sevrinsky}, {Jao}, {Henry},
  {Riedel}, {Subasavage}, {Lurie}, {Ianna}, \& {Finch}}]{Winters(2017)}
{Winters}, J.~G., {Sevrinsky}, R.~A., {Jao}, W.-C., {et~al.} 2017, \aj, 153, 14

\bibitem[{{Winters} {et~al.}(2018){Winters}, {Irwin}, {Newton}, {Charbonneau},
  {Latham}, {Han}, {Muirhead}, {Berlind}, {Calkins}, \&
  {Esquerdo}}]{Winters(2018)}
{Winters}, J.~G., {Irwin}, J., {Newton}, E.~R., {et~al.} 2018, \aj, 155, 125

\bibitem[{{Winters} {et~al.}(2019{\natexlab{a}}){Winters}, {Henry}, {Jao},
  {Subasavage}, {Chatelain}, {Slatten}, {Riedel}, {Silverstein}, \&
  {Payne}}]{Winters(2019a)}
{Winters}, J.~G., {Henry}, T.~J., {Jao}, W.-C., {et~al.} 2019{\natexlab{a}},
  \aj, 157, 216

\bibitem[{{Winters} {et~al.}(2019{\natexlab{b}}){Winters}, {Medina}, {Irwin},
  {Charbonneau}, {Astudillo-Defru}, {Horch}, {Eastman}, {Halley Vrijmoet},
  {Henry}, {Diamond-Lowe}, {Winston}, {Barclay}, {Bonfils}, {Ricker},
  {Vanderspek}, {Latham}, {Seager}, {Winn}, {Jenkins}, {Udry}, {Twicken},
  {Teske}, {Tenenbaum}, {Pepe}, {Murgas}, {Muirhead}, {Mink}, {Lovis},
  {Levine}, {L{\'e}pine}, {Jao}, {Henze}, {Fur{\'e}sz}, {Forveille},
  {Figueira}, {Esquerdo}, {Dressing}, {D{\'\i}az}, {Delfosse}, {Burke},
  {Bouchy}, {Berlind}, \& {Almenara}}]{Winters(2019b)}
{Winters}, J.~G., {Medina}, A.~A., {Irwin}, J.~M., {et~al.} 2019{\natexlab{b}},
  \aj, 158, 152

\bibitem[{{Winters} {et~al.}(2020){Winters}, {Irwin}, {Charbonneau}, {Latham},
  {Medina}, {Mink}, {Esquerdo}, {Berlind}, {Calkins}, \&
  {Berta-Thompson}}]{Winters(2020a)}
{Winters}, J.~G., {Irwin}, J.~M., {Charbonneau}, D., {et~al.} 2020, \aj, 159,
  290

\bibitem[{{Woitas} {et~al.}(2003){Woitas}, {Tamazian}, {Docobo}, \&
  {Leinert}}]{Woitas(2003)}
{Woitas}, J., {Tamazian}, V.~S., {Docobo}, J.~A., \& {Leinert}, C. 2003, \aap,
  406, 293

\bibitem[{{Ziegler} {et~al.}(2018){Ziegler}, {Law}, {Baranec}, {Morton},
  {Riddle}, {De Lee}, {Huber}, {Mahadevan}, \& {Pepper}}]{Ziegler(2018)}
{Ziegler}, C., {Law}, N.~M., {Baranec}, C., {et~al.} 2018, \aj, 156, 259

\end{thebibliography}

\clearpage


\end{document}